\documentclass[aps,pre,reprint,groupedaddress,superscriptaddress]{revtex4-1}

\usepackage{graphicx}
\usepackage{amssymb}
\usepackage[tbtags]{amsmath}
\usepackage{bm}
\usepackage{hyperref}
\usepackage{lipsum}
\usepackage{bbold}
\usepackage{comment}
\usepackage[utf8]{inputenc}
\usepackage[capitalize]{cleveref}

\begin{document}

\title{Sliding-dependent electronic structures of alternating-twist tetralayer graphene}

\author{Kyungjin Shin}
\thanks{These authors contributed equally to this work.}
\affiliation{Department of Physics and Astronomy, Seoul National University, Seoul 08826, Korea}

\author{Jiseon Shin}
\thanks{These authors contributed equally to this work.}
\affiliation{Department of Physics, University of Seoul, Seoul 02504, Korea}
\affiliation{CTO Division, LG Electronics, Seocho R\&D Campus, Seoul 06772, Korea}

\author{Yoonsung Lee}
\affiliation{Department of Physics, University of Seoul, Seoul 02504, Korea}

\author{Hongki Min}
\email[Contact author: hmin@snu.ac.kr]{}
\affiliation{Department of Physics and Astronomy, Seoul National University, Seoul 08826, Korea}

\author{Jeil Jung}
\email[Contact author: jeiljung@uos.ac.kr]{}
\affiliation{Department of Physics, University of Seoul, Seoul 02504, Korea}
\affiliation{Department of Smart Cities, University of Seoul, Seoul 02504, Korea}


\begin{abstract}
We study the electronic structure of alternating-twist tetralayer graphene, especially near its magic angle $\theta = 1.75^\circ$,
for different AA, AB, and SP sliding geometries at their middle interface that divides two twisted bilayer graphenes.
This sliding dependence is shown for the bandwidths, band gaps, and $K$-valley Chern numbers of the lowest-energy valence and conduction bands as a function of twist angle and interlayer potential difference.
Our analysis reveals that the AA sliding is most favorable for narrow bands and gaps, and the AB sliding is most prone to developing finite valley Chern numbers.
We further analyze the linear longitudinal optical absorptions as a function of photon energy and the absorption map in the moir\'{e} Brillouin zone for specific transition energies. 
A self-consistent Hartree calculation reveals that the AA system's electronic structure is the most sensitive to variations in carrier density.
\end{abstract}


\maketitle

\section{Introduction}

The discovery of ordered phases in twisted bilayer graphene (T2G)~\cite{Cao2018a, Cao2018b} ignited vibrant research into moir\'{e} two-dimensional (2D) material properties.
Many other conceivable combinations beyond T2G studied since then
include twisted double bilayer graphene (T2BG)~\cite{Chebrolu2019, Koshino2019, Burg2019,Lee2019,Choi2019, Liu2020},
twisted triple bilayer graphene~\cite{JShin2022nearly},
twisted mono-bilayer graphene (TMBG)~\cite{Morell2013, Marton2020,  Xu2021,Chen2021,Polshyn2020,Park2020, Lei2021, Zewen2019,Xiao2019,Carr2020,MA202118},
alternating-twist multilayer graphene~\cite{Park2021,Hao2021,Cao2021,Park2022,Burg2022,Zhang2022, khalaf2019magic,KShin2023,JShin2021,MA202118,Xiao2019, Carr2020, Lei2021, Zewen2019, Lopez-Bezanilla2020,Phong2021,Dumitru2021,Fang2021,Yu2023,Kolar2023},
and consecutive-twist trilayer graphene~\cite{Mora2019,Zhu2020,Zhang2021,Mao2023,Devakul2023,Popov2023,Guerci2023,Nakatsuji2023}, among many possible systems.
Up to now, the alternating-twist multilayer graphene systems from bilayer to pentalayer samples have shown attributes of robust superconductivity such as the zero-resistivity dome in the phase diagram, the nonlinear $I$-$V$ characteristics, and the Fraunhofer-like Josephson phase coherence~\cite{Cao2018a,Park2021, Hao2021,Cao2021,Park2022,Burg2022,Zhang2022}, 
while all those features have not been observed simultaneously in other structures such as T2BG~\cite{Liu2020} and TMBG~\cite{Xu2021}.
Hence, an in-depth study of the expected electronic structure of alternating-twist multilayer graphene is of current interest. 

We study the non-interacting single-particle electronic structure of alternating-twist tetralayer graphene (AT4G) by presenting the bandwidths, band gaps, and topological and optical properties
and evaluate the impact of the self-consistent Hartree term for various band filling factors for three interlayer slidings of the middle interface.
It is of interest to quantify the impact of electrostatic Hartree screening as a function of carrier density since it was already shown that it tends to smoothen the pronounced charge localization at the AA stacking sites in the case of T2G~\cite{Louk2019,Goodwin2020,Guinea2018,Cea2019,Cea2020,Cea2022,Novelli2020,Ding2022,Lewandowski2021,Choi2021,Zhang2020,Ming2020,Ming2021,Wei2023}.
\begin{figure*}
\begin{center}
\includegraphics[width=1\textwidth]{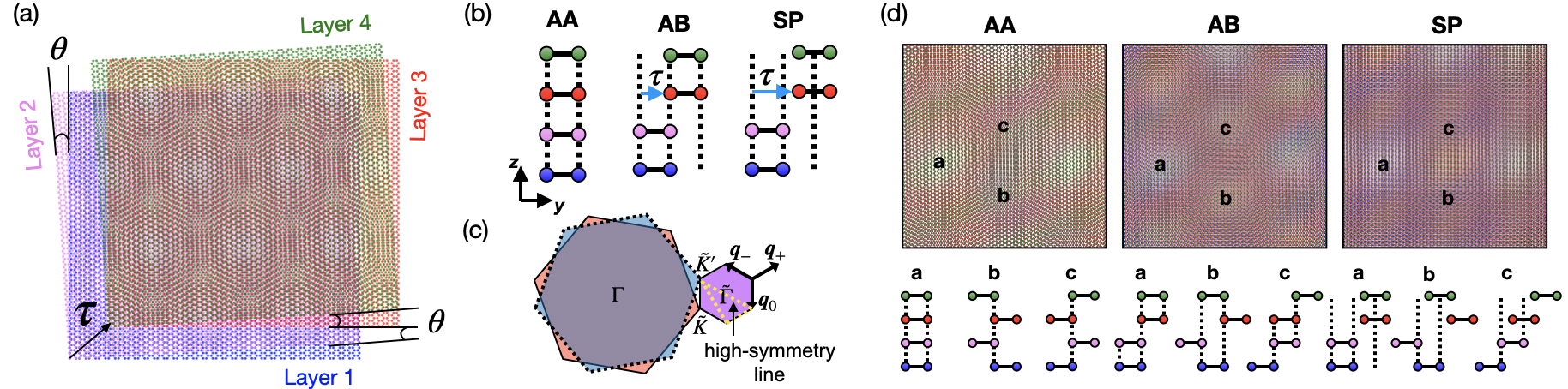}\\
\end{center}
\caption{
(a) Moir\'{e} pattern of AT4G at $\theta = 1.75^\circ$ with an arbitrary sliding $\bm{\tau}$ of the upper twisted bilayer. 
(b) Schematic diagrams for the atomic configurations with selective starting stackings, AA, AB, and SP at the middle moir\'{e} interface in the $yz$-plane. 
(c) Schematic diagram for BZs of the four monolayers of graphene (the red hexagon encompassed by the solid line and the blue shaded one with the dotted line) and the associated mBZ (the smaller purple hexagon). We indicate the high-symmetry line by the yellow dashed line for the one-dimensional band structure of Fig.~\ref{fig2}. 
(d) Moir\'{e} patterns generated from the three starting stackings AA, AB, and SP at $\theta = 1.75^\circ$, and the schematic diagrams for their local commensurate stackings, denoted by $\textbf{a}$, $\textbf{b}$, and $\textbf{c}$.
}\label{fig1}
\end{figure*}
This paper is organized as follows.
In Sec.~\ref{Model}, we describe our model Hamiltonian explicitly.
In Sec.~\ref{R1}, we explain the electronic structures of AT4G in the single-particle framework at the magic angle $\theta = 1.75^\circ$ in particular and provide an in-depth description by presenting the band gaps and bandwidths in terms of how the electronic properties change with different twist angles, starting stackings, and displacement fields.
In Sec.~\ref{R2}, we sweep the parameter space of displacement fields and twist angles in search of parameter sets for strong effects of electron correlations by estimating the ratio of the Coulomb interactions to the bandwidth.
We show the phase diagram of the single-valley Chern numbers of the lowest-energy electron and hole bands as a function of the displacement fields and the twist angles in Sec.~\ref{R3}.
We present the longitudinal linear optical absorption for different displacement fields as a function of the photon energy, and the optical transition map in the moir\'{e} Brillouin zone (mBZ) for selected transition energies in Sec.~\ref{R4}.
We give the quasiparticle band structures modified by the self-consistent Hartree potential for the three different starting stackings tuned by sliding the upper twisted bilayer for different filling factors from $-4$ to $+4$ in Sec.~\ref{R5}.
We summarize our work in Sec.~\ref{Summary}.

\section{Model}\label{Model}

We study AT4G tetralayer graphene in which the successive layers have relative alternating-twist angles $\pm \theta$ in the broad range of angles $0 \leq \theta \leq 2.5^\circ$. We account for a global shift of the upper twisted bilayer by a displacement $\bm{\tau} = (\tau_x, \tau_y)$, as shown in Fig.~\ref{fig1}(a).
We pay special attention to the three cases of layer shift of the upper twisted bilayer, AA ($\tau_y = 0$), AB ($\tau_y = a/\sqrt{3}$), and the saddle point (SP; $\tau_y = a\sqrt{3}/2$) for $\tau_x = 0$, as shown in Fig.~\ref{fig1}(b), and leave the detailed electronic structures for a global shift to Appendix A.
In Fig.~\ref{fig1}(d) we present the moir\'{e} patterns for the three cases where the starting stackings are AA, AB, and SP for the middle moir\'{e} interface
and show on the bottom the structures of the commensurate local stackings $\textbf{a}$, $\textbf{b}$, and $\textbf{c}$.
For the AA systems, we have a triangular lattice of $\textbf{a}$ bright spot sites with maximally coincident local AA stacking regions between all layers.
The local stacking $\textbf{a}$ of the AB systems also forms a triangular shape but exhibits a different local density of states,
while for the SP systems a stripe pattern emerges along the $y$-direction.

In Fig.~\ref{fig1}(c), the Brillouin zones (BZs) of the four individual graphene layers are illustrated by the red ($-\theta/2$) and the blue ($+\theta/2$) larger hexagons, and the small pink hexagon indicates the mBZ of our model, whose corners match each corner of the two original BZs.
Throughout this paper, we conduct our numerical calculations for a single spin valley.

\begin{figure*}
\begin{center}
\includegraphics[width=1\textwidth]{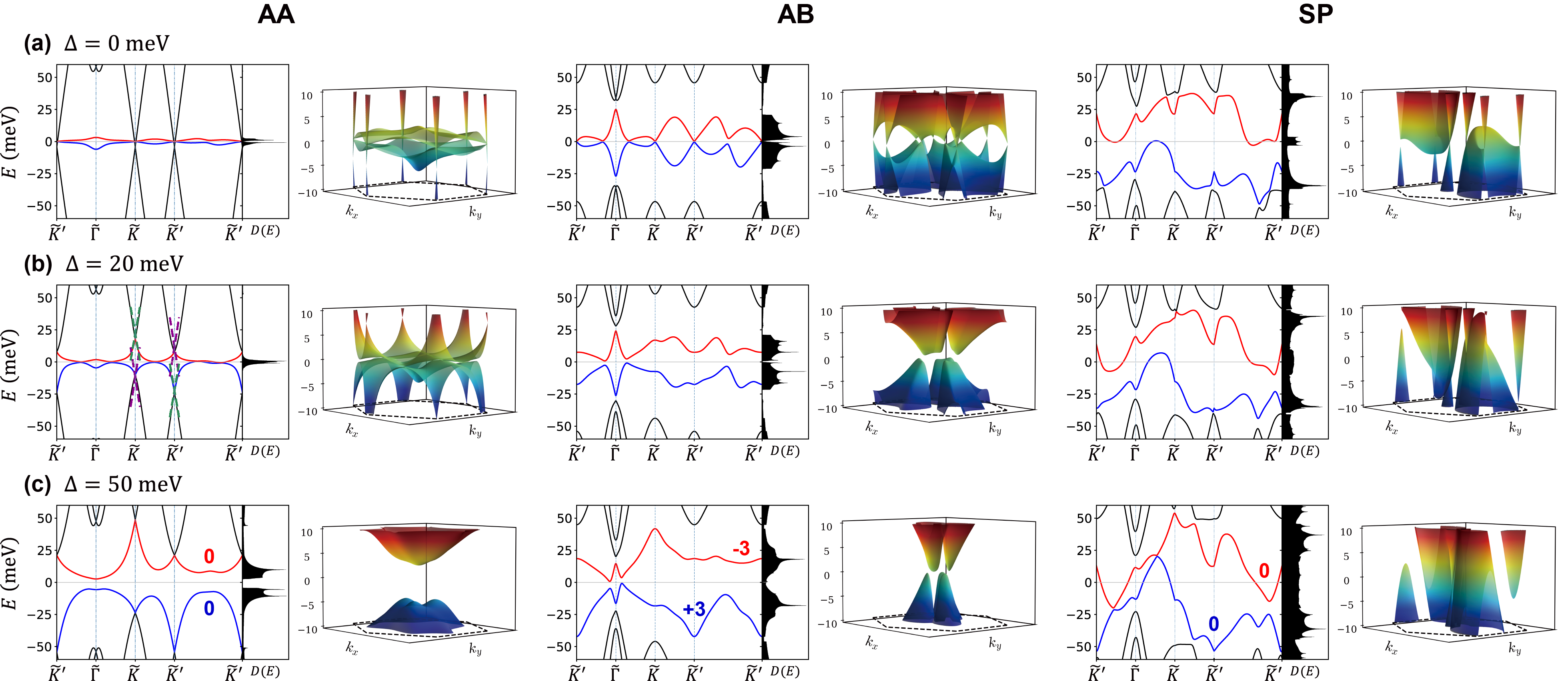}\\
\end{center}
\caption{
Single-particle electronic band structures for the alternating-twist tetralayer graphene at $\theta = 1.75^\circ$ with three different shifts of the upper twisted bilayer, AA (left), AB (middle), and SP (right), for three different electrical gatings: (a) $\Delta = 0$~meV, (b) $20$~meV, and (c) $50$~meV. For each case, we show the band structures along the high-symmetry line ($\tilde{K'}$-$\tilde{\Gamma}$-$\tilde{K}$-$\tilde{K'}$-$\tilde{K'}$) on the left together with their densities of states in the middle and the three-dimensional energy dispersion in the mBZ in the right column. For $\Delta = 50$~meV, the $K$-valley Chern numbers for the valence (blue) and conduction (red) bands are $0$ for AA sliding; $+3$ and $-3$, respectively, for AB-sliding; and $0$ for SP sliding.
}\label{fig2}
\end{figure*}

The continuum model Hamiltonian of AT4G projected to the $K$-valley is given by
\begin{equation}
H_\textrm{AT4G} (\theta) = \left(\begin{array}{cccc}
h_1^- & T_1(\boldsymbol{r}) & 0 & 0  \\
T_1^{\dagger}(\boldsymbol{r}) & h_2^+ & T_2(\boldsymbol{r}) & 0\\
0 & T_2^{\dagger}(\boldsymbol{r}) & h_3^-  &T_3(\boldsymbol{r})\\
0 & 0 & T_3^{\dagger}(\boldsymbol{r}) & h_4^+  \\
\end{array}\right).
\end{equation}
The diagonal blocks are $h^\pm_i = D^\dagger (\pm \theta/2) ~ h_i ~ D (\pm \theta/2)$, where $h_i(\theta = 0) = v_{\rm F} \boldsymbol{p} \cdot \boldsymbol{\sigma}$ are two-dimensional Dirac cones on the $i$th layer ($i = 1, \cdots, 4$), with $v_{\rm F} = \sqrt{3} \vert t_0 \vert a/ 2\hbar$, $t_0 = -3.1$ eV, and the lattice constant $a = 2.46 ~\textrm{\AA}$.
The general form of a spin rotation operator is defined as $D(\phi) = e^{-i \phi \sigma_z /2}$ when the $z$ axis is the rotation axis and $\sigma_z$ is the $z$-component Pauli matrix.
Regarding the electrical gating on the system, we add a matrix of the form $V = \textrm{diag} (\bar{V}_1, ~\bar{V}_2, ~\bar{V}_3, ~\bar{V}_4)$, where $\bar{V}_i = V_i \mathbb{1}$ are $2\times2$ matrices introduced to apply a displacement field on the $i$th layer in our model.
We assume that the gating voltage $V_i$ drops monotonically such that $V_i$ satisfies $V_1 = 3\Delta/2$, $V_2 = \Delta/2$, $V_3 = -V_2$, and $V_4 = -V_1$, implying that the potential differences between consecutive layers are $\Delta$.

The interlayer couplings $T_k(\boldsymbol{r})$ at a moir\'{e} interface between the $k$th and ($k+1$)th layers can be defined as~\cite{Jung2014,Bistritzer2010b}
\begin{equation}
T_k(\boldsymbol{r}) = \sum_{j=0, \pm} e^{-i m_k \boldsymbol{q}_j \cdot \boldsymbol{r}} T^j_{l, l'},
\end{equation}
where $m_k = (-1)^k$, $\boldsymbol{q}_0 = k_{\theta} (0, -1)$, and $\boldsymbol{q}_\pm = k_{\theta} (\pm \sqrt{3}/2, 1/2)$, as illustrated in Fig.~\ref{fig1}(c)
with $k_{\theta}=2k_{\rm D}\sin(\theta/2)$, a moir\'{e} version of the Dirac momentum $k_{\rm D} = 4\pi/3 a$.
The components $T^j_{l, l'}$ are given in terms of the relative stacking configurations at each moir\'{e} interface.
Provided that the location of the reference atom of the upper (lower) layer is indicated by $\bm{\tau}_{l'}$ ($\bm{\tau}_{l}$), the matrix can be expressed in the form $T^j_{l, l'} = \omega_{l,l'} \exp(i j \bm{G}_j \cdot \bm{\tau})$, where $\bm{\tau} = (\tau_x, ~\tau_y)= \bm{\tau}_l - \bm{\tau}_{l'}$, $\bm{G}_0 = (0, 0)$, and $\bm{G}_\pm = k_{\rm D} (-3/2, \pm \sqrt{3}/2)$.
In this work, we assume out-of-plane lattice relaxation taking unequal interlayer tunneling parameters with $\omega_{\rm{AA'}} = \omega_{\rm{BB'}} = 0.0939$~eV and $\omega_{\rm{AB'}} = \omega_{\rm{BA'}} = 0.12$~eV for intra- and inter-sublattice hoppings respectively as used in Ref.~\cite{Chebrolu2019}.
We can also implement a global shift of the upper twisted bilayer by changing $\bm{\tau}_{l'}$ continuously at the middle moir\'{e} interface between the second and third layers as illustrated in Fig.~\ref{fig1}(a).
See Appendix A for the detailed numerical results of continuously shifted upper twisted bilayer.

%

\section{Single-particle electronic structures}
\label{R1}

We first show the single-particle electronic band structures of AT4G at the twist angle $\theta = 1.75^\circ$ for three cases where the second and third layers are AA, AB, or SP stacked, distinguishing the four layers of AT4G as two sets of  T2Gs under an interlayer potential difference $\Delta = 0$, $20$, and $50$~meV, as presented in Figs.~\ref{fig2}(a-c).
Here, we set the zero-energy level at the charge neutrality point (CNP).
For each panel, we juxtapose a band structure along the high-symmetry line ($\tilde{K^\prime}$-$\tilde{\Gamma}$-$\tilde{K}$-$\tilde{K^\prime}$-$\tilde{K^\prime}$) as illustrated in Fig.~\ref{fig1}(c), its density of states (DOS), and the associated three-dimensional energy dispersion in the mBZ.

In the case of AA sliding without an electric field, the Hamiltonian is decomposed into two T2G Hamiltonians with different interlayer couplings~\cite{khalaf2019magic,KShin2023}, which results in the formation of two distinct Dirac cones at mBZ corners.
At the magic angle, one of the two Dirac cones becomes flat; thus, its low-energy band structure is characterized by the flattest bands with linear Dirac bands at these points.
After application of the electric field, the two Dirac cones with different velocities are hybridized and split, as shown in the leftmost panel of Fig.~\ref{fig2}(b).
When $\Delta$ is $20$~meV, the Dirac points are split by about $28.35$ and $31.56$~meV at $\tilde{K}$ and $\tilde{K^\prime}$, respectively, which is close to the analytic result obtained from first-order degenerate-state perturbation theory~\cite{KShin2023}, represented by the dashed lines.

As the interlayer potential difference becomes much larger than the bandwidth, the perturbation approach is no longer valid, and the two lowest-energy bands are detached, as shown in the leftmost panel of Fig.~\ref{fig2}(c).
The primary gap $\delta_p$, which refers to the energy difference between the two lowest-energy bands near the CNP, opens when the electric field is larger than $\Delta \sim 25$~meV, and $\delta_{p} \sim 8$~meV when $\Delta = 50$~meV.
This distinct tunable gap emerges due to the shift of the Dirac points and the flattening of the two lowest bands, which is eventually closed when the next higher and lower energy bands approach each other near $\tilde{\Gamma}$.
Moreover, one can see in the three-dimensional energy dispersion that the lowest-energy valence band has three humps in the vicinity of the $\tilde{\Gamma}$ point, reflecting the $C_{3z}$ symmetry.
Note that in the case of AA-stacked AT3G, $\delta_{p}$ does not open even under a larger $\Delta \sim 100$~meV~\cite{JShin2021} due to the appearance of an additional decoupled Dirac cone in the absence of the potential.


\begin{figure*}
\begin{center}
\includegraphics[width=1\textwidth]{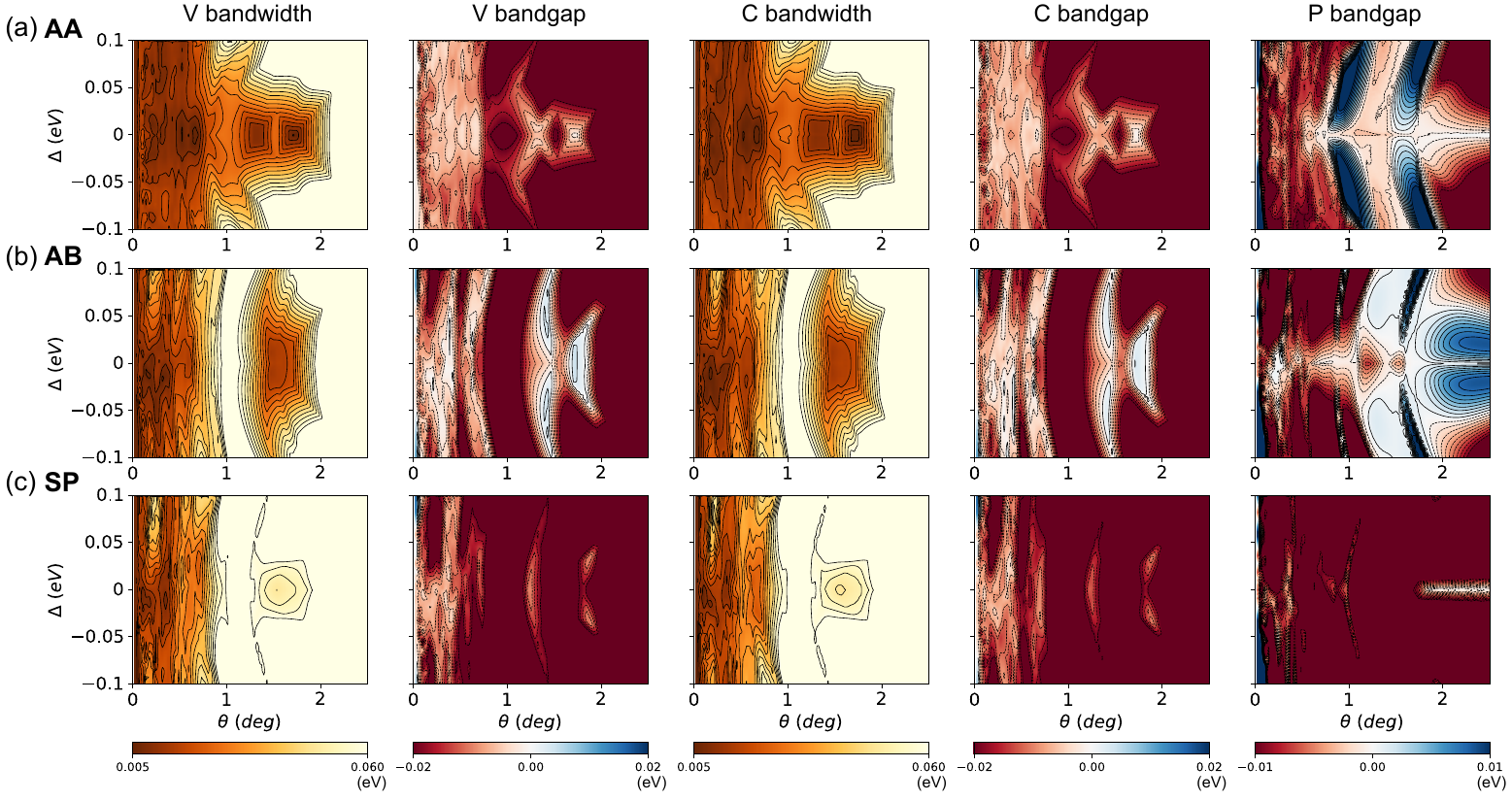}\\
\end{center}
\caption{
Valence (V) bandwidth, V secondary band gap, conduction (C) bandwidth, C secondary band gap, and primary (P) band gap of AT4G as a function of the interlayer potential difference $\Delta$ and the twist angle $\theta$ with the upper twisted bilayer slid by (a) $0$, (b) $a/\sqrt{3}$, and (c) $a\sqrt{3}/2$ in the $y$ direction, leading to AA, AB, and SP starting stackings, respectively, at the middle moir\'{e} interface.
}\label{fig3}
\end{figure*}


When the upper twisted bilayer is slid by $a/\sqrt{3}$ in the $y$ direction, the starting stacking at the middle moir\'{e} interface is AB, and the associated band structures for the three interlayer potential differences are shown in the middle columns of Fig.~\ref{fig2}.
The lowest-energy bands form a loop like shape along the one-dimensional high-symmetry path, giving almost zero DOS at the CNP, and open the secondary gaps $\delta_s$, which are defined as energy differences between the lowest-energy conduction (valence) band and the next higher (lower) energy band.
Once the interlayer potential difference is turned on, the Dirac points at $\tilde{K}$ and $\tilde{K}^{\prime}$ at the CNP disappear, and $\delta_p$ is still open, while both $\delta_s$ are closed.
On the other hand, for the SP case, the bandwidth remains significantly larger than in other commensurate cases regardless of the interlayer potential difference,
and both $\delta_p$ and $\delta_s$ are closed for all three values of the interlayer potential difference.

We present the bandwidth and $\delta_s$ of the lowest-energy valence and conduction bands and the primary band gap in a row in Figs.~\ref{fig3}(a-c) for the AA-, AB-, and SP-stacked upper twisted bilayer as a function of the interlayer potential difference $\Delta$ and the twist angle $\theta$.
(See Appendix A for the bandwidths and band gaps of the lowest-energy bands for a global shift $\boldsymbol{\tau}$ at twist angle $\theta = 1.75^\circ$.)
It is noteworthy that the diagrams of $\Delta$ versus $\theta$ are symmetric with respect to the line of the zero interlayer potential difference for all cases,
implying that the direction of the interlayer potential difference makes no difference because our system is composed of two identical T2Gs.
Furthermore, the diagrams for the lowest-energy valence and conduction bands are mostly the same in the whole parameter space.

For the case where the upper and the lower T2Gs are AA stacked, particularly narrow bandwidths appear at twist angles $\theta = 1.4^\circ$ and $1.75^\circ$ for $\theta > 1^\circ$, as shown in Fig.~\ref{fig3}(a), whereas the bandwidth decreases significantly for smaller twist angles $\theta \leq 1^\circ$.
Interestingly, it turns out that an additional first magic angle exists in AT4G around $\theta = 0.67^\circ$ according to the magic angle hierarchy~\cite{khalaf2019magic}, and we provide the band structures for the twist angles in the vicinity of $\theta = 1.4^\circ$ and $\theta = 0.67^\circ$ in Appendix B.
Both $\delta_s$ for the lowest-energy bands are closed in the entire parameter space since the Dirac cone-shaped next-higher-energy bands are attached to the lowest-energy bands.
$\delta_p$ becomes finite around $\theta = 1^\circ$ and $1.75^\circ$ for finite interlayer potential differences.
See Appendix B for the electronic band structure with twist angle $\theta = 1^\circ$ for several interlayer potential differences.

As shown in Fig.~\ref{fig3}(b), when two T2Gs have AB sliding, the lowest-energy bands become narrow around $\theta = 1.4^\circ$ for a small interlayer potential difference.
Furthermore, both $\delta_s$ are open at $\theta = 1.4^\circ$ up to a large electric field $\vert \Delta \vert \sim 100$~meV, and $\theta = 1.75^\circ$ for a smaller electric field less than around $50$~meV.
$\delta_p$ is open for angles larger than $2^\circ$ for an interlayer potential difference $\vert \Delta \vert \lesssim 50$~meV and at $1.75^\circ$ up to a large electric field $\vert \Delta \vert \sim 100$~meV.
We provide the electronic band structure at $\theta = 2.3^\circ$ as an example for several interlayer potential differences in Appendix B.

If the upper T2G is SP stacked we cannot flatten and isolate the lowest-energy bands in the entire parameter space considered in Fig.~\ref{fig3}(c).


\begin{figure*}
\begin{center}
\includegraphics[width=1\textwidth]{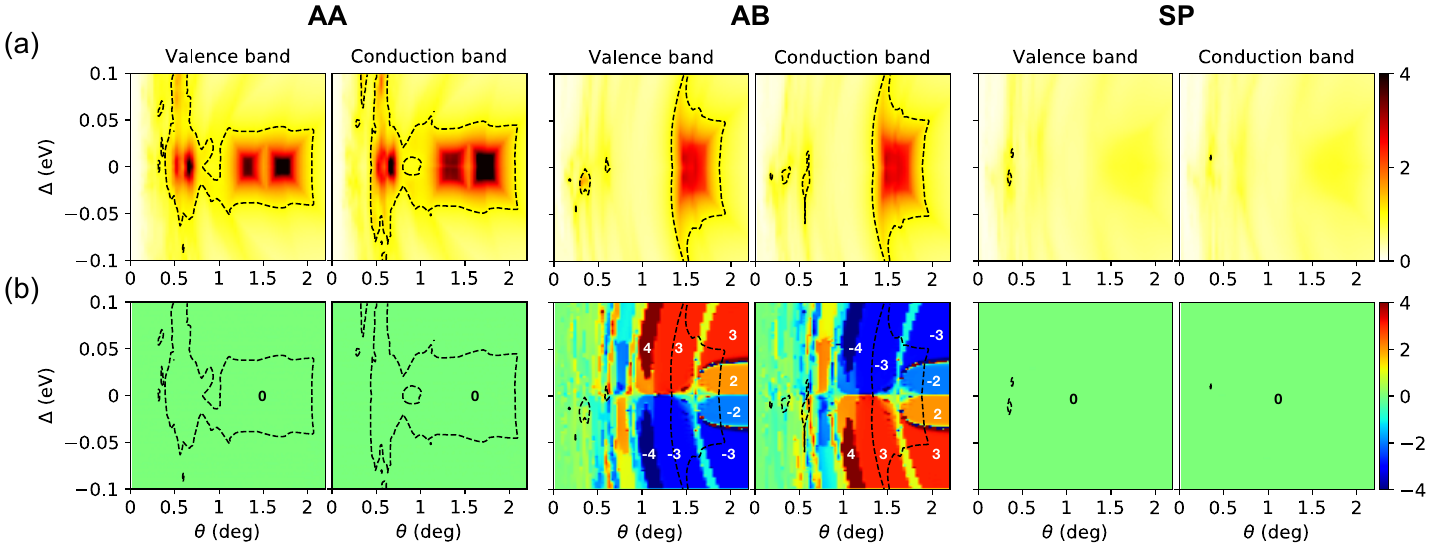}\\
\end{center}
\caption{
(a) Ratio of the Coulomb potential to the bandwidth $U/W$ and (b) $K$-valley Chern number phase diagrams as a function of the interlayer potential difference $\Delta$ and the twist angle $\theta$ for the lowest-energy valence and conduction bands when the two sets of T2G are AA stacked (left two columns), AB stacked (middle two columns), and SP stacked (right two columns). Here, the contour denoted by the black dashed lines corresponds to $U/W = 1$.
}\label{fig4}
\end{figure*}


\section{Coulomb interaction versus bandwidth}\label{R2}

We estimate the ratio of the Coulomb interaction $U$ to the bandwidth $W$ utilizing the general form of the Coulomb potential $U = e^{2}/(\epsilon_{r} l_{M})$, where the moir\'{e} length is $l_M \sim a/\theta$ and the relative dielectric constant is $\epsilon_r = 4$~\cite{Jung2013}. The ratio $U/W > 1$ represents the regime of strong correlations due to greater Coulomb energy than the kinetic energy that is associated with the band flatness.

Figure~\ref{fig4}(a) shows $U/W$ as a function of the interlayer potential difference $\Delta$ versus twist angle $\theta$.
For the case where the two T2Gs are AA stacked (left two columns), the ratio $U/W$ becomes significantly large at $\theta = 0.67^\circ$, $1.4^\circ$, and $1.75^\circ$ (see Appendix B for $\theta = 0.67^\circ$ and $1.4^\circ$ band structures)
and in the vicinity of these angles in the presence of a small interlayer potential difference in both the valence and conduction bands.
On the other hand, for the AB-stacked T2Gs, the ratio $U/W$ is larger than 1 for a broad range of angles $\theta = 1.5^\circ$ and $2^\circ$.
For the SP case, there is no relevant parameter set for a high ratio of the Coulomb interaction to the bandwidth in the entire parameter space.
Importantly, in the region with $\theta \sim 1.75^{\circ}$ and low interlayer bias, the AA-stacked T2Gs exhibit a pronounced ratio of $U/W$ in both the valence and conduction bands,
suggesting that significant electron correlation is expected for AA systems compared to other configurations such as AB and SP with these parameter sets.

\section{Chern numbers}\label{R3}

In Fig.~\ref{fig4}(b), we evaluate the Chern numbers for the lowest-energy valence and conduction bands in the parameter space of interlayer potential differences $\Delta$ and twist angles $\theta$.
We superpose the contour corresponding to $U/W = 1$ with dashed lines to identify the valley Chern numbers of the lowest-energy valence and conduction bands that remain nearly flat. 
In this work, we consider only the $K$ valley for the Chern number, and we define the Chern number of the $n$th energy band as

\begin{equation}
C_n = \frac{1}{2 \pi} \int_\textrm{mBZ} d^2 \boldsymbol{k} ~ \Omega_n (\boldsymbol{k}),
\end{equation}
where the Berry curvature $\Omega_n (\boldsymbol{k})$ is given as \cite{Xiao2010}
\begin{equation}
\Omega_n (\boldsymbol{k}) = -2 \sum_{n' \neq n} \textrm{Im} \Bigg[ \frac{\langle n \vert \frac{\partial H}{\partial k_x} \vert n' \rangle \langle n' \vert \frac{\partial H}{\partial k_y} \vert n \rangle}{(E_{n'} - E_n)^2} \Bigg].
\label{Bcurv}
\end{equation}

If two T2Gs are AA stacked, the two lowest-energy bands are topologically trivial in the entire parameter space, while the AB-stacked T2Gs have rich phase diagrams with well-defined nontrivial Chern numbers.
The diagrams for the valence and conduction bands have the same absolute values but opposite signs for the Chern numbers.
Moreover, the change in the direction of the interlayer potential difference reverses the sign.
In particular, the Chern number around $\theta \sim 1^\circ$ for the valence (conduction) band is $+4$ ($-4$) for a positive $\Delta$ and $-4$ ($+4$) for a negative $\Delta$ when $\vert \Delta \vert \gtrsim 25$~meV, while $+2$ ($-2$) for a positive $\Delta$ and $-2$ ($+2$) for a negative $\Delta$ when $\vert \Delta \vert \lesssim 25$~meV.
The diagrams for both the valence and conduction bands are largely divided into two parts by an arched narrow area from $\Delta = -0.1$~eV to $\Delta = +0.1$~eV centered around $\theta \approx 1.6^\circ$, which has a band Chern number of $+1$ ($-1$) for a positive $\Delta$ in the valence (conduction) band, and $-1$ ($+1$) for a negative $\Delta$ in the valence (conduction) band.
The Chern number on the left side of the arched stripe area is $+3$ ($-3$) for a positive $\Delta$ in the valence (conduction) band, and the sign of the Chern number is reversed for a negative $\Delta$.
The right side of the stripe area is again split into two areas whose Chern numbers are $+3$ ($-3$) for $\Delta \gtrsim 50$~meV and $+2$ ($-2$) for $\Delta \lesssim 50$~meV in the valence (conduction) band, with the reverse sign for a negative $\Delta$.
For the case of the SP-stacked T2Gs, the two lowest-energy bands are topologically trivial, like in the AA-stacked case.

\section{Optical properties}\label{R4}
\begin{figure*}
\begin{center}
\includegraphics[width=0.9\textwidth]{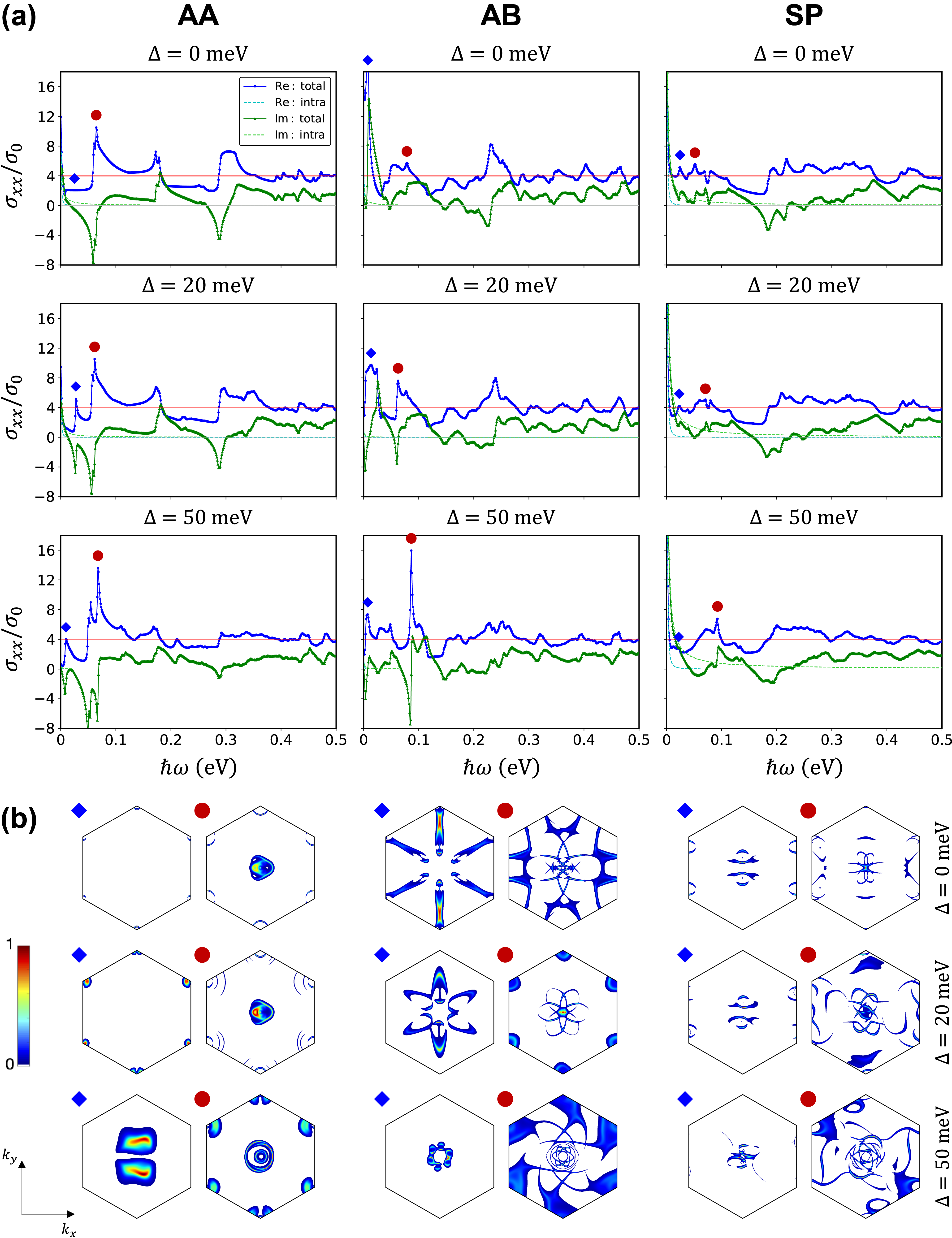}\\
\end{center}
\caption{
(a) Real and imaginary parts of the linear longitudinal optical absorptions denoted by the blue line with circles and the green line with triangles, respectively, for the three differently stacked T2Gs, AA (left), AB (middle), and SP (right),
under three different interlayer potential differences, $\Delta = 0$, $20$, and $50$~meV. The red solid line in each panel represents the constant $4\sigma_{0}$, which stands for the optical absorption value of the four layers of graphene without a twist.
(b) Maps of the real part of the optical absorption in the mBZ at specific transition energies of particularly high contributions indicated by the blue diamonds and the red circles in (a). 
}\label{fig5}
\end{figure*}

We first calculate the linear longitudinal optical conductivity $\sigma_{xx}(\omega)$, which is derived from  linear response theory as

\begin{equation}\label{Eq_Kubo}
\begin{split}
\sigma_{xx}(\omega)
&=- \frac{ie^2}{\hbar} \int\frac{d^2 k}{(2\pi)^2} \sum_{n, m}
\frac{f_{n\bm{k}} - f_{m\bm{k}}}{\varepsilon_{n\bm{k}} - \varepsilon_{m\bm{k}}} \\
& \times \frac{\vert \langle n,\bm{k} \vert \hbar\hat{v}_{x} \vert m,\bm{k} \rangle \vert^2}{\hbar\omega + \varepsilon_{n\bm{k}} - \varepsilon_{m\bm{k}} + i\eta},
\end{split}
\end{equation}
where $f_{n\bm{k}}=1/[1+e^{(\varepsilon_{n\bm{k}}-\mu)/k_{\rm B}T}]$ is the Fermi-Dirac distribution function,
and $\hat{v}_{x}=\partial \hat{H} /  \hbar \partial  k_{x}$ is the velocity operator along the $x$ direction.
The eigenstate and corresponding eigenvalue of the $n$th energy band at $\bm{k}$ in the mBZ are, respectively, denoted as $\vert n,\bm{k} \rangle$ and $\varepsilon_{n\bm{k}}$ with an infinitesimal positive constant $\eta$.
For numerical calculations, we employ a finite broadening term set to $\eta=1$~meV.
The real part of the linear longitudinal optical conductivity accounts for optical transitions upon the incidence of light at frequency $\omega$.
We plot the real part of the optical conductivity as a function of the frequency of incident light with the blue lines with circles in Fig.~\ref{fig5}(a) and with the two-dimensional color map in the mBZ in Fig.~\ref{fig5}(b).
On the other hand, the imaginary part of the optical conductivity, which represents the delayed current due to the driving field according to linear response theory~\cite{Stauber2013}, is depicted with the green lines with triangles in Fig.~\ref{fig5}(a).
In Fig.~\ref{fig5}(a), the dashed lines correspond to the Drude term due to intraband transitions.

Figure~\ref{fig5}(a) shows the real and imaginary parts of the linear longitudinal optical conductivity $\sigma_{xx}(\omega)$, normalized by the optical conductivity of monolayer graphene $\sigma_{0}=g_{\rm sv}e^{2}/16\hbar$ with a spin-valley degeneracy factor $g_{\rm sv}=4$, as a function of the photon energy $\hbar \omega$ in eV.
We consider a system with a zero chemical potential for three different starting stackings, AA, AB, and SP, at the middle moir\'{e} interface (row) and for the different interlayer potential differences (column), $\Delta = 0$, $20$, and $50$~meV.
The system of four layers of graphene exhibits $4\sigma_{0}$ in the high-frequency limit~\cite{Min2009}, and we indicate this value with the red solid line in all of the panels in Fig.~\ref{fig5}(a).
Note that the real parts of optical absorption in all panels show a fluctuating behavior around the constant value $4 \sigma_0$ for a large photon energy limit.

When the upper T2G is placed on the lower one with AA stacking $\Delta=0$~meV, the real part of optical absorption has a constant value $2 \sigma_0$ for small transition energy at $\hbar \omega \lesssim 50$~meV, and we pick one transition energy (denoted by the blue diamond marker at $\hbar \omega \sim 25$~meV) to see the contribution in the mBZ.
Furthermore, it has a prominent peak around $\hbar \omega \sim 60$~meV, which we denote with the red circle.
More peaks appear for larger transition energy, but we focus our attention on photon energy less than $100$~meV.
As the bandwidth of the lowest valence and conduction bands is smaller than $25$ meV, the optical conductivity at the blue diamond does not arise from interband transitions between those bands.
Instead, the constant optical conductivity $2\sigma_{0}$ originates from interband transitions between the linear Dirac bands, depicted as annular-shaped areas near $\tilde{K}$ and $\tilde{K}^\prime$ in the hexagon marked by the blue diamond in the top left of Fig.~\ref{fig5}(b).
For the hexagon marked with the red circles, the contribution arises mostly from the $\tilde{\Gamma}$ point as well as annular-shaped areas near mBZ corners.

If the interlayer potential difference is set to $\Delta = 20$~meV, a peak appears at $\hbar \omega \sim 25$~meV, indicated with the blue diamond, and $\tilde{K}$, $\tilde{K}^{\prime}$ and their vicinity are responsible for this peak.
The existing peak around $\hbar \omega \sim 60$~meV under the zero interlayer potential difference remains, and the $\tilde{\Gamma}$ point and its neighborhood mainly contribute to this peak, and the two concentric thin annular-shaped areas centered at $\tilde{K}$ and $\tilde{K}^{\prime}$ play a small part.

If the interlayer potential difference increases up to $\Delta=50$~meV, the band structure possesses a finite primary gap $\delta_{p} \sim 8$~meV, and the low-frequency conductivity converges to zero.
At $\hbar\omega \sim 10$~meV, slightly above $\delta_{p}$, a new peak appears, indicated by the blue diamond, due to the optical transitions near $\tilde{\Gamma}$.
The red circle indicates the optical conductivity at $\hbar\omega \sim 67.5$~meV, exhibiting additional optical transitions at moir\'{e} $\tilde{K}$ and $\tilde{K}^{\prime}$ compared to the case of the blue diamond.

When the upper T2G is stacked on the lower one with AB stacking, we find one prominent peak in the real part of the optical absorption at a small transition energy $\hbar \omega \sim 7$~meV, which is indicated by the blue diamond.
This peak is mainly contributed by the six identical paths along the high-symmetric line from $\tilde{\Gamma}$ to $\tilde{K}$ and $\tilde{K}^{\prime}$.
Concerning the next peak indicated by the red circle, the contribution is from broad areas across the mBZ.

When $\Delta$ increases up to $20$~meV, we find two distinct peaks in the real part of the conductivity around $\hbar \omega \sim 25$ and $64$~meV, denoted by the blue diamond and red circle.
The optical transition at the blue diamond is contributed from the narrowly connected broad range adjacent to $\tilde{K}$ and $\tilde{K}^{\prime}$ as well as the hexagonally distorted annular area near $\tilde{\Gamma}$.
The transition at the red circle originates from the circular area near $\tilde{K}$ and $\tilde{K}^{\prime}$ and snow flake like pieces near $\tilde{\Gamma}$.

For $\Delta = 50$~meV, the real part of the conductivity fluctuates near $4\sigma_{0}$ with a small peak at $\hbar\omega \sim 50$~meV denoted by the blue diamond, and for larger transition energy it has a significant peak at $\hbar\omega \sim 89$~meV, denoted by the red circle.
The optical transitions at both the blue diamond and red circle are centrosymmetric with respect to the origin of the mBZ.
The transitions at the red circle occur in almost the entire range across the mBZ with a pinwheel shape.

The optical transitions that occur when one sheds light on the SP-stacked T2Gs neither surpass $8\sigma_{0}$ nor deviate too much from $4\sigma_{0}$.
We also sample two peaks for a transition energy less than $0.1$~eV, denoted by the blue diamond and the red circle, and show contributions in the momentum-space map.

Importantly, the optical transitions partially reveal the symmetries of the system, and the absorption maps in Fig.~\ref{fig5}(b) reflect these symmetries.
The $C_{2x}$ symmetric picture is observed when $\Delta = 0$, while this symmetry is broken when $\Delta \neq 0$ because it flips the layer sequence.
This symmetry breaking is pronounced at mBZ corners and is marginal near $\tilde{\Gamma}$.

\section{self-consistent Hartree potential}\label{R5}

The effects of the electrostatic Hartree potential on electronic structures of T2G were studied extensively in Refs.~\cite{Louk2019,Goodwin2020,Guinea2018,Cea2019,Cea2020,Cea2022,Novelli2020,Ding2022,Lewandowski2021,Choi2021,Ming2020,Ming2021,Wei2023,Zhang2020}. Commonly reported features include the smoothening of charge distribution, which gives rise to a significant change in the lowest-energy bands.
The Fermi level approximately pins to Van Hove singularities (VHS) upon changing the filling factor, retaining the gapless Dirac cones at $\tilde{K}$ and $\tilde{K}^{\prime}$, and nearly preserving the gap size at $\tilde{\Gamma}$.
The fact that similar effects occur in the band structures of AT3G has been corroborated~\cite{Phong2021, Dumitru2021,Fang2021,Yu2023}.


\begin{figure*}
\begin{center}
\includegraphics[width=1.0\textwidth]{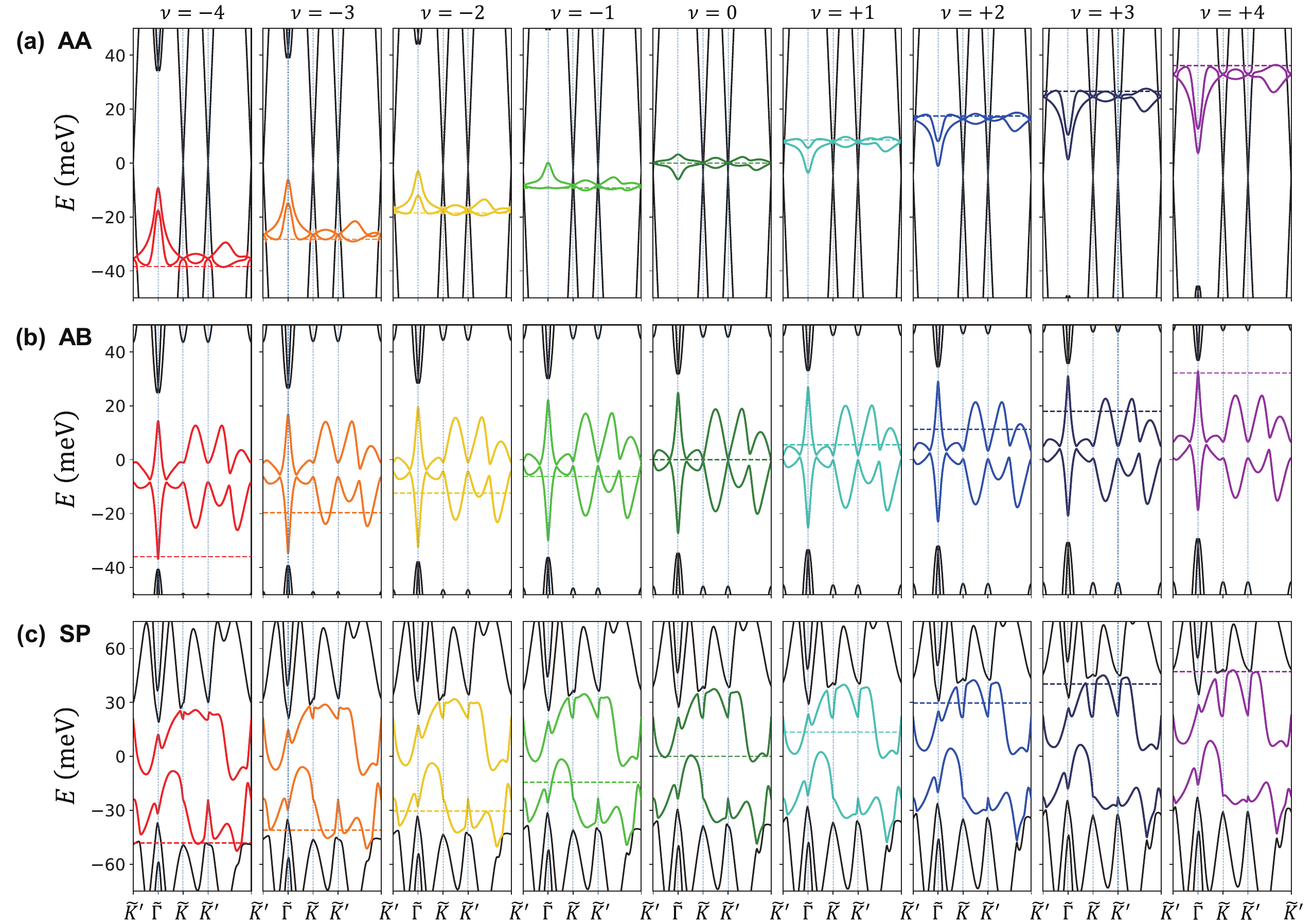}\\
\end{center}
\caption{
Band structures obtained by the self-consistent Hartree method with $\epsilon_{r}=30$ when two T2Gs with twist angle $\theta = 1.75^\circ$ are stacked as (a) AA, (b) AB, and (c) SP for different filling factors $\nu = -4$ to $4$. The colored solid lines corresponding to different filling factors represent the low-energy bands, with the corresponding Fermi level indicated by a dashed line in the same color.
}\label{fig6}
\end{figure*}


%
In this work, we employ the Hartree potential, described in more detail in Appendix C, confined to a single valley given the time reversal symmetry between the two valleys, $K$ and $K^{\prime}$.
The dielectric constant, a control parameter for the strength of interaction, was set within various ranges in prior studies;
however, using the typical values $\epsilon_{r}=4$--$6$ leads to discrepancies between theoretical and experimental results because these values significantly overestimate the Hartree potential~\cite{Choi2021}.
For this reason, we choose a dielectric constant $\epsilon_r = 30$, assuming weaker interactions due to screening~\cite{Ming2020,Ming2021,Wei2023}, and leave the results with $\epsilon_r = 4$ for Appendix C.
On the other hand, the dielectric constant for the out-of-plane interaction due to the layer charge difference is maintained as $\epsilon_{\perp}=4$ throughout this paper~\cite{Kolar2023}.
The band structures obtained with the self-consistent Hartree method for cases where two T2Gs with $\theta = 1.75^\circ$ are stacked as AA, AB, and SP are plotted in Figs.~\ref{fig6}(a-c), respectively.

For the AA-stacked T2Gs, like for T2G and AT3G, the Dirac points shift upward (downward) relative to the $\tilde{\Gamma}$ point as electrons (holes) are doped, with the Fermi level pinning to the VHS, 
while the gap at $\tilde{\Gamma}$ remains nearly constant across filling factors ranging from $-4$ to $4$.
In contrast, the effect of the Hartree potential is nearly vanishing for the AB- and SP-stacked T2Gs compared to the AA configuration.
In T2G, the Hartree potential redistributes the inhomogeneous charge distribution, which is mainly due to the localization of flat bands at the AA spots~\cite{Laissardiere2010,Uchida2014,Koshino2018,YChoi2019}.
Our Hartree band structures indicate that the charge distribution in the AA-stacked T2Gs is highly localized at the $\textbf{a}$ region in Fig.~\ref{fig1}(d) compared to other configurations, which might result from its narrower bandwidth.
Furthermore, the low-energy bands in Fig.~\ref{fig6}(a) become flatter near the Fermi level, which could potentially lead to an enhancement of the effective interaction or correlation effects.

\section{SUMMARY}\label{Summary}

We investigated continuum model electronic structures for AT4G, which can be thought of as two vertically stacked equal twist angle T2Gs.
We attempted to provide an overall map of expected low-energy bandwidths as a function of twist angle, interlayer potential difference, and sliding between layers.
This class of system remains of high interest because alternating-twist multilayer graphene provides a simple and robust material platform with which we can realize moir\'{e} superconductivity.
The increase in number of layers in alternating-twist multilayer graphene allows achieving larger magic angles and therefore an enhanced mechanical stability that is less susceptible to moir\'{e} strains.
Moreover, a shorter moir\'{e} length scale associated with a larger magic angle potentially enhances the magnitude of the Coulomb interaction strength in the moir\'{e} supercell and therefore the critical temperatures for the ordered states.

The AT4G system has some peculiarities such as the possibility of opening a band gap upon application of a vertical electric field, 
while structurally it is only slightly more complex than the trilayer.
Unlike a single T2G that results in the same moir\'{e} patterns and physical properties regardless of the sliding between layers, alternating-twist multilayer graphene consisting of more than two layers has completely different electronic structures depending on the relative shift between the moir\'{e} patterns formed by contiguous layers.
The three twisted interfaces have three moir\'{e} patterns that can slide relative to each other, exhibiting different low-energy band structures depending on the sliding configurations.
For our study, we chose in particular three sliding geometries of the upper twisted bilayer against the lower one, where the moir\'{e} patterns formed by the two T2Gs are AA, AB, and SP stacked.

In our work, we provided maps of the bandwidth and other electronic structure details in the large parameter space of twist angle $\theta$ versus the interlayer potential difference $\Delta$ and gave in-depth descriptions for the first magic angle, $\theta = 1.75^\circ$.
We showed that AA-stacked AT4G has a wide primary gap between topologically trivial lowest-energy bands with the help of the interlayer potential difference $\Delta \gtrsim 25$~meV, 
while AB-stacked AT4G has nearly zero DOS at the CNP with secondary gaps with Chern numbers of $\pm 3$ for the lowest valence and conduction bands.
The bandwidth of AA-stacked AT4G is significantly smaller than that of the other two configurations, resulting in a relatively higher ratio of the Coulomb interaction strength versus bandwidth.
We also provide the band structures for selected cases of ($\theta$, $\Delta$) in Appendix B,
and we discussed the real and imaginary parts of optical transitions for AA-, AB-, and SP-stacked AT4G with interlayer potential differences of $0$, $20$, and $50$~meV, which can be directly measured in experiments.
Our study shows that the electron correlation effect is the most important for AT4G with AA-stacked moir\'{e} patterns, consisting of two T2Gs stacked on top of each other
given the relative favoring of the narrow bands and gaps, while the AB sliding is the most prone to developing finite valley Chern numbers.
The charge redistribution makes the lowest-energy bands significantly asymmetric due to Hartree screening, while this effect is largely suppressed for AB and SP slidings.
Our work confirms that the trade-off between system complexity and the expected robustness of the electronic structure properties makes AT4G an interesting system to explore the correlation-driven physical phenomena beyond T2G and AT3G graphene moir\'{e} flat band systems.

\begin{acknowledgments}
K.S. and H.M. were supported by the National Research Foundation of Korea (NRF) grants funded by the Korea government (MSIT) (Grant No. 2023R1A2C1005996); the Creative-Pioneering Researchers Program through Seoul National University (SNU); and the Center for Theoretical Physics.
J.S. acknowledges support from NRF (Grant No. 2021R1A6A3A01087281).
J.J. acknowledges support from the Basic Study and Interdisciplinary R\&D Foundation Fund of the University of Seoul (2022).
\end{acknowledgments}

\begin{figure*}
\begin{center}
\includegraphics[width=1\textwidth]{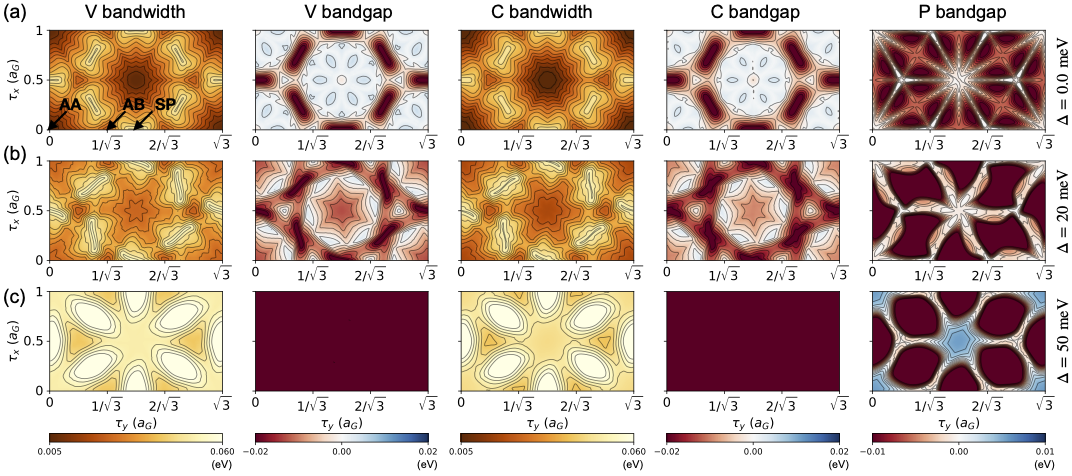}\\
\end{center}
\caption{ 
Bandwidths and secondary and primary band gaps of the lowest-energy valence and conduction bands of AT4G at twist angle $\theta = 1.75^\circ$ for the three different displacement fields, (a) $\Delta = 0$, (b) $20$, and (c) $50$ meV, as a function of the global shift $\bm{\tau} = (\tau_x, \tau_y)$ in units of the graphene lattice constant $a$.
}\label{Afig1}
\end{figure*}

\section*{Appendix A. Lowest-energy bands with respect to global shift}\label{AppendixA}

To get a full-scale understanding of shifting the upper twisted bilayer in AT4G, we conducted a numerical analysis for a general shift in the range of $0$ to $1$ for $\bm{\tau}_x$ and $0$ to $\sqrt{3}$ for $\bm{\tau}_y$ in units of $a$, as shown in Fig.~\ref{Afig1}.
Figure~\ref{Afig1} helps the reader readily compare the bandwidths and band gap sizes of different sliding geometries directly with the color and contours.
The AA, AB, and SP slidings are denoted by the arrows in the leftmost panel in the first column.
Starting from the left, each column describes the bandwidth, secondary band gaps, and the primary (P) band gaps of the lowest-energy valence (V) and conduction (C) bands.
Each row has a assigned for different interlayer potential differences, $0$, $20$, and $50$ meV.
Note that the valence and conduction bands have nearly the same diagrams. 

Without the displacement field, AA sliding has the narrowest bandwidth.
However, as the interlayer potential difference increases to $\vert \Delta \vert \gtrsim 25$~meV, the AB case exhibits the narrowest bandwidth, while the SP bandwidth remains the largest for both the valence and conduction bands.
Regarding the secondary gap, AA and SP do not have a finite gap, while AB always has a nearly zero gap of $2$--$3$~meV, which gradually decreases as the interlayer potential difference of $\vert \Delta \vert \gtrsim 20$~meV is applied.
In AA, the gap closes for $\Delta=0$~meV, but it opens broadly for large interlayer potential difference. 

\begin{figure*}
\begin{center}
\includegraphics[width=0.8\textwidth]{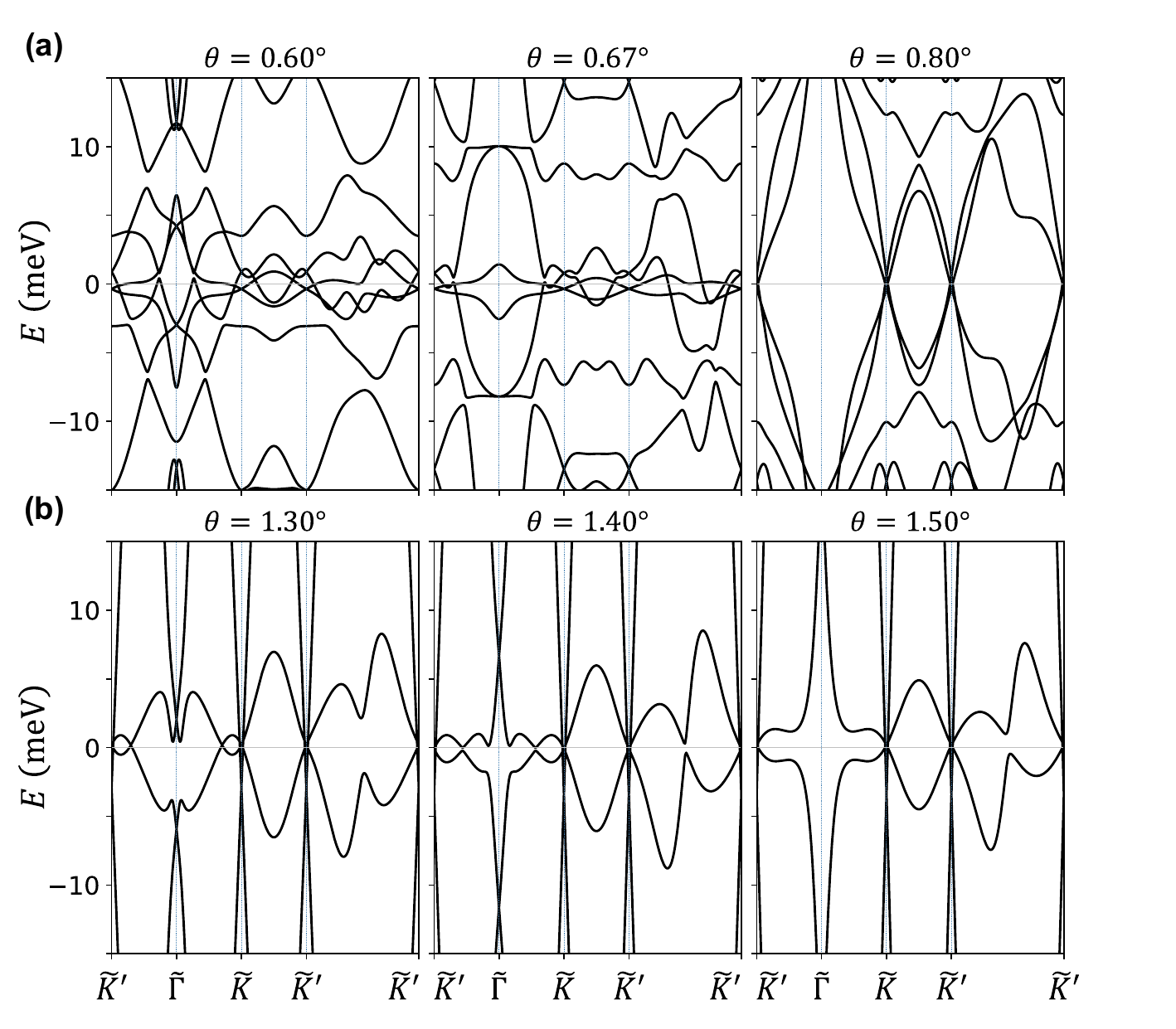}\\
\end{center}
\caption{ 
Single-particle band structures of AT4G with the AA starting stacking at the middle moir\'{e} interface without a displacement field for different twist angles: (a) $\theta = 0.6^\circ$, $0.67^\circ$, and $0.8^\circ$ and (b) $\theta = 1.3^\circ$, $1.4^\circ$, and $1.5^\circ$.
}\label{Afig2_1}
\end{figure*}

\begin{figure*}
\begin{center}
\includegraphics[width=0.8\textwidth]{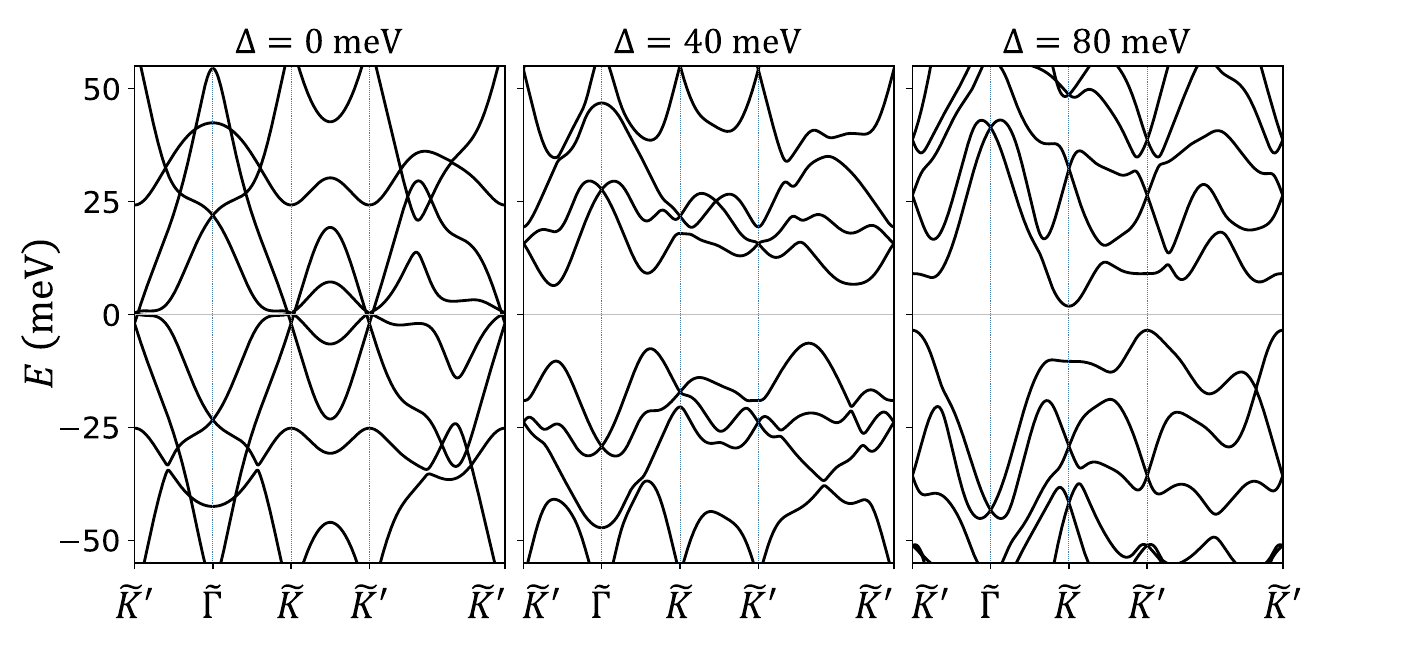}\\
\end{center}
\caption{ 
Single-particle band structures of AT4G with the AA starting stacking at the middle moir\'{e} interface at $\theta = 1^\circ$ for different displacement fields, $\Delta = 0$, $40$, and $80$ meV.
}\label{Afig2_2}
\end{figure*}

\begin{figure*}
\begin{center}
\includegraphics[width=0.8\textwidth]{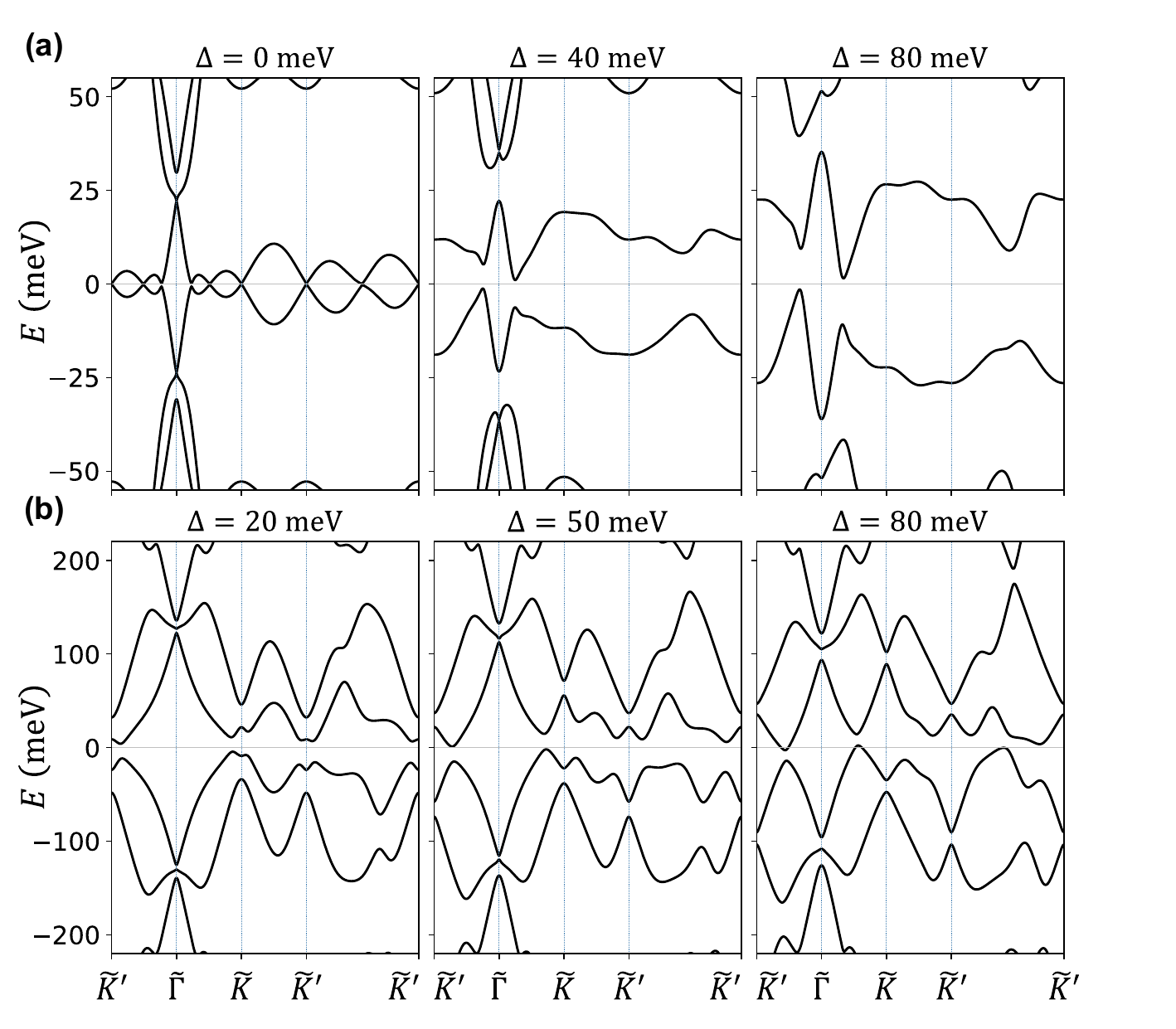}\\
\end{center}
\caption{ 
Single-particle band structures of AT4G with the AB starting stacking at the middle moir\'{e} interface (a) at twist angle $\theta = 1.4^\circ$ for different displacement fields, $\Delta = 0$, $40$, and $80$ meV, and (b) at twist angle $\theta = 2.3^\circ$ for different displacement fields, $\Delta = 20$, $50$, and $80$ meV.
}\label{Afig2_3}
\end{figure*}


\section*{Appendix B. Band structures for selected angles}\label{AppendixB}

We provide single-particle band structures for several selected cases in which the bandwidth is very small or $\delta_s$ and $\delta_p$ are widely open.
We first show in Fig.~\ref{Afig2_1}(a) the band structures for the AA sliding without a displacement field for three twist angles, $\theta = 0.6^\circ$, $0.67^\circ$, and $0.8^\circ$,
to see the changes in the electronic structure as the twist angle varies in the vicinity of the other first magic angle of AT4G ($0.67^\circ$) according to the magic angle hierarchy~\cite{khalaf2019magic}.
One can find several sets of nearly particle-hole symmetric bands overlaid on one another, and there are two low-energy bands similar to those of T2G decoupled from the others.
The T2G-like low-energy band becomes flat with a bandwidth less than $5$~meV near the first magic angle $\theta = 0.67^\circ$, and the bandwidth increases rapidly up to $\sim 1^\circ$.

According to Fig.~\ref{fig3}(a), there is another twist angle at which the bandwidths of both the conduction and valence bands are suppressed around $\theta = 1.4^\circ$ for the middle moir\'{e} interface with the AA starting stacking.
Figure~\ref{Afig2_1}(b) shows the band structures without the displacement field for three angles, $\theta = 1.3^\circ$, $1.4^\circ$, and $1.5^\circ$.
The lowest-energy band is attached to the higher-energy bands at the $\tilde{\Gamma}$ point, and the bandwidths of the lowest-energy band grow as the twist angle increases up to $\sim 1.5^\circ$.

Figure~\ref{Afig2_2} shows the single-particle band structures of AT4G with the AA sliding at a twist angle of $1^\circ$ where $\delta_p$ is broadly open for finite displacement field, as shown in Fig.~\ref{fig3}(a).
We provide the energy bands for three different displacement fields, $\Delta = 0$, $40$, and $80$ meV.
$\delta_p$ is open after the field is turned on, while both $\delta_s$ are closed in all ranges of the parameter space. 

In Fig.~\ref{fig3}(b), where the second layer is stacked on the third layer with the AB starting stacking, one can see at $\theta = 1.4^\circ$ that $\delta_s$ is open for a finite displacement field for both the conduction and valence bands.
We choose three values of the displacement field, $\Delta = 0$, $40$, and $80$ meV, and show the low-energy bands in Fig.~\ref{Afig2_3}(a).
For all three cases, $\delta_p$ remains nearly zero, but $\delta_s$ is open for a finite displacement field. 

In the broad range satisfying $\vert \Delta \vert \lesssim 50$ meV and $\theta \gtrsim 2^\circ$, $\delta_p$ is open, but $\delta_s$ for the valence and conduction bands are closed.
We present the band structures for the cases with displacement fields $\Delta = 20$, $50$, and $80$ meV with the AB sliding and a twist angle of $\theta = 2.3^\circ$ in Fig.~\ref{Afig2_3}(b).


\section*{Appendix C. Hartree potential}\label{AppendixC}

\begin{figure*}
\begin{center}
\includegraphics[width=1.0\textwidth]{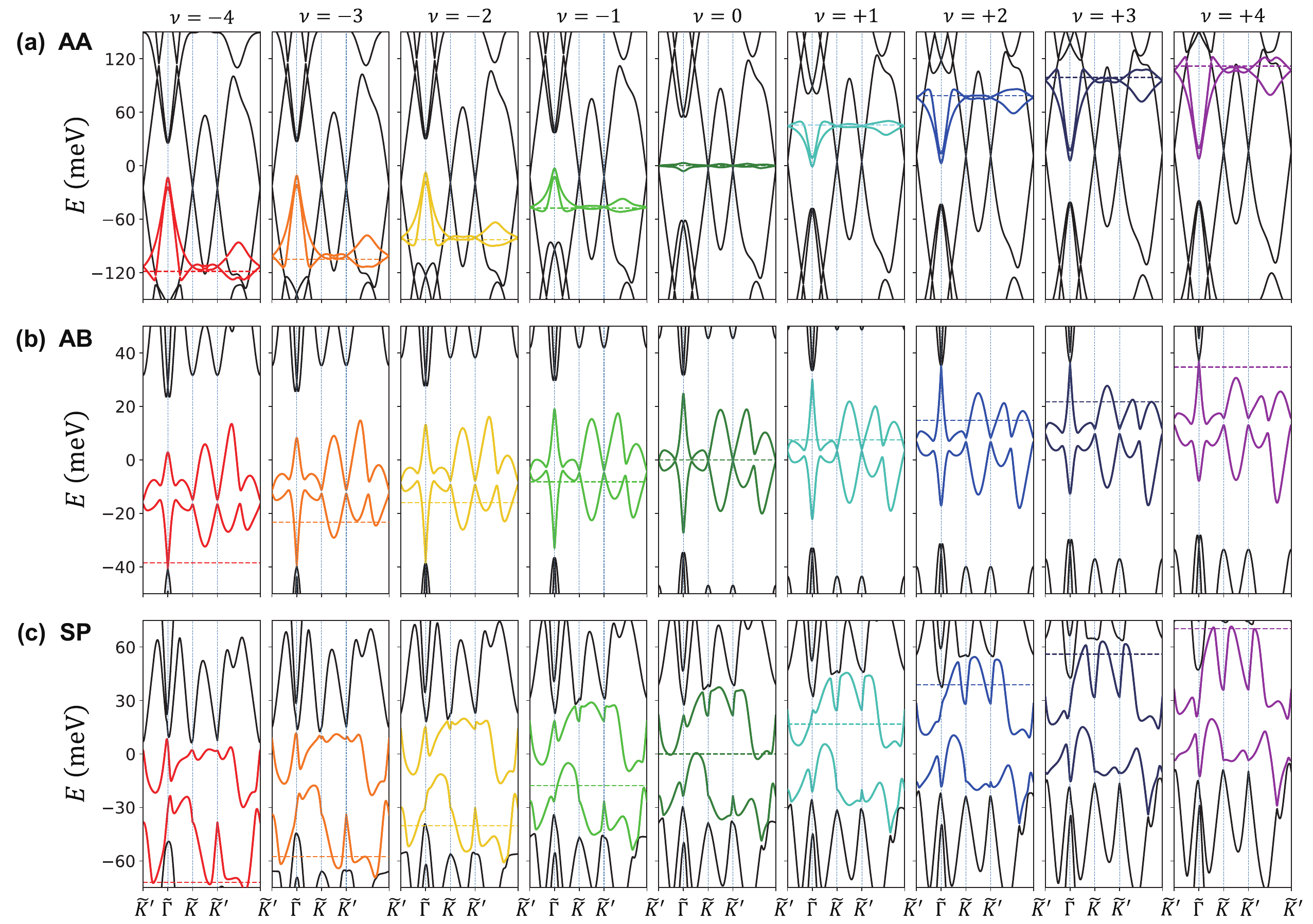}\\
\end{center}
\caption{ 
Same as Fig.~\ref{fig6} for $\epsilon_{r}=4$.
}\label{Afig3}
\end{figure*}

%
In this appendix, we provide an explicit description of the self-consistent Hartree potential used in this work.
The total Hamiltonian can largely be split into two parts: the non-interacting Hamiltonian $H_{0}$, which was dealt with in the main text, and the Hartree potential $V_{H}$.
The classical electrostatic interaction between electrons can be treated as a single-particle operator by employing mean-field Hartree theory.
The Hartree equation consists of in-plane and out-of-plane terms~\cite{Kolar2023}.
The in-plane term is given as
follows~\cite{Phong2021,Guinea2018, Cea2019,Cea2020,Cea2022,Novelli2020,Ding2022,Lewandowski2021,Choi2021}:
\begin{equation}
V_{H}^{\rm in}(\bm{r}) = \frac{1}{A_{M}} \sum_{\bm{G} \neq 0} V(\bm{G}) \rho_{\bm{G}} e^{ i \bm{G} \cdot \bm{r}},
\label{Eq_apC_1}
\end{equation}
whereas the out-of-plane term corresponds to the Poisson equation, resulting in the electrostatic potential between the $k$th and $(k+1)$th layers as
\begin{equation}
V_{k,k+1}^{\rm out} = \frac{2\pi e^2 d}{ \epsilon_{\perp}} \left(-\sum_{\ell=1}^{k} n_{\ell}+\sum_{\ell=k+1}^{4} n_{\ell} \right),
\end{equation}
with the interlayer separation $d=3.5~\textrm{\AA}$ and the electron density for the $\ell$th layer denoted as $n_{\ell}$.
Here, $A_{M}=l_{M}^{2}\sqrt{3}/2$ is the area of the moir\'{e} unit cell, $V(\bm{G})$ is the Fourier-transformed Coulomb potential, and $\rho_{\bm{G}}$ is the Fourier component of the electron density corresponding to the wave vector $\bm{G}$, defined by
\begin{equation}
\rho_{\bm{G}} = g_{\rm sv} A_{M} \int \frac{d^2 k}{(2\pi)^2} \sum_{m, \bm{G}'}
\phi_{m,\bm{G}'}^{\dagger}(\bm{k})\phi_{m,\bm{G}+\bm{G}'}(\bm{k}),
\end{equation}
where the band index $m$ runs over the occupied states and $\phi_{m,\bm{G}}(\bm{k})$ denotes the electron amplitude for a state with momentum $\bm{k}+\bm{G}$, defined relative to the CNP.
We consider only the first shell in the reciprocal space, which is closest to the center of the mBZ, as its contributions are dominant compared to the higher-order harmonics~\cite{Guinea2018}.

For the configuration of our system, we assume a dual-gate setup consisting of two metallic gates sandwiching two substrates, with the AT4G sample placed between them, leading to the 2D Coulomb interaction given by
\begin{equation}
V(\bm{q}) = \frac{2\pi e^2}{ \epsilon_r \vert \bm{q} \vert} \tanh(\vert \bm{q} \vert D_g),
\end{equation}
where $D_g = 10$~nm is the gate separation determined by the thickness of the substrate~\cite{Bernevig2021, Dumitru2021,Ming2021,Goodwin2020}.
The additional factor $\tanh(\vert \bm{q} \vert D_g)$ originates from the screening by the infinitely generated imaginary charges reflected by the metallic gates~\cite{Dumitru2021,Kolar2023, Bernevig2021}, assuming that the sample is in contact with the source and drain at each horizontal end.

Figures~\ref{Afig3}(a-c) show the band structures obtained with the self-consistent Hartree method with $\epsilon_r=4$ for AA-, AB-, and SP-stacked AT4G at $\theta=1.75^\circ$.
Compared to the result with $\epsilon_{r}=30$ in Fig.~\ref{fig6}, the low-energy bandwidth of the AA-stacked AT4G with $\epsilon_{r}=4$ widens, reaching about $100$~meV at full filling ($\vert \nu \vert=4$), as illustrated in Fig.~\ref{Afig3}(a).
This substantial broadening of the bandwidth may result from the overestimation of the Hartree contribution while neglecting the exchange and correlation effects~\cite{Choi2021}.
Despite these differences in band broadening, the band structures retain several features, such as the shifting of Dirac points relative to the $\tilde{\Gamma}$ point and a nearly invariant gap at $\tilde{\Gamma}$.


\begin{thebibliography}{0}%
\makeatletter
\providecommand \@ifxundefined [1]{%
 \@ifx{#1\undefined}
}%
\providecommand \@ifnum [1]{%
 \ifnum #1\expandafter \@firstoftwo
 \else \expandafter \@secondoftwo
 \fi
}%
\providecommand \@ifx [1]{%
 \ifx #1\expandafter \@firstoftwo
 \else \expandafter \@secondoftwo
 \fi
}%
\providecommand \natexlab [1]{#1}%
\providecommand \enquote  [1]{``#1''}%
\providecommand \bibnamefont  [1]{#1}%
\providecommand \bibfnamefont [1]{#1}%
\providecommand \citenamefont [1]{#1}%
\providecommand \href@noop [0]{\@secondoftwo}%
\providecommand \href [0]{\begingroup \@sanitize@url \@href}%
\providecommand \@href[1]{\@@startlink{#1}\@@href}%
\providecommand \@@href[1]{\endgroup#1\@@endlink}%
\providecommand \@sanitize@url [0]{\catcode `\\12\catcode `\$12\catcode
  `\&12\catcode `\#12\catcode `\^12\catcode `\_12\catcode `\%12\relax}%
\providecommand \@@startlink[1]{}%
\providecommand \@@endlink[0]{}%
\providecommand \url  [0]{\begingroup\@sanitize@url \@url }%
\providecommand \@url [1]{\endgroup\@href {#1}{\urlprefix }}%
\providecommand \urlprefix  [0]{URL }%
\providecommand \Eprint [0]{\href }%
\providecommand \doibase [0]{http://dx.doi.org/}%
\providecommand \selectlanguage [0]{\@gobble}%
\providecommand \bibinfo  [0]{\@secondoftwo}%
\providecommand \bibfield  [0]{\@secondoftwo}%
\providecommand \translation [1]{[#1]}%
\providecommand \BibitemOpen [0]{}%
\providecommand \bibitemStop [0]{}%
\providecommand \bibitemNoStop [0]{.\EOS\space}%
\providecommand \EOS [0]{\spacefactor3000\relax}%
\providecommand \BibitemShut  [1]{\csname bibitem#1\endcsname}%
\let\auto@bib@innerbib\@empty
\end{thebibliography}%


\begin{thebibliography}{999}

\bibitem{Cao2018a}
Y. Cao, V. Fatemi, S. Fang, K. Watanabe, T. Taniguchi, E. Kaxiras, and P. Jarillo-Herrero,
Unconventional superconductivity in magic-angle graphene superlattices, 
Nature (London) {\bf 556}, 43 (2018).

\bibitem{Cao2018b}
Y. Cao, V. Fatemi, A. Demir, S. Fang, S. L. Tomarken, J. Y. Luo, J. D. Sanchez-Yamagishi, K. Watanabe, T. Taniguchi, E. Kaxiras, R. C. Ashoori, and P. Jarillo-Herrero,
Correlated insulator behaviour at half-filling in magic-angle graphene superlattices, 
Nature (London) {\bf 556}, 80 (2018).

\bibitem{Chebrolu2019}
N. R. Chebrolu, B. L. Chittari, and J. Jung,
Flat bands in twisted double bilayer graphene,
Phys. Rev. B {\bf 99}, 235417 (2019).

\bibitem{Koshino2019}
M. Koshino,
Band structure and topological properties of twisted double bilayer graphene,
Phys. Rev. B {\bf 99}, 235406 (2019).

\bibitem{Burg2019}
G. W. Burg, J. Zhu, T. Taniguchi, K. Watanabe, A. H. MacDonald, and E. Tutuc,
Correlated Insulating States in Twisted Double Bilayer Graphene,
Phys. Rev. Lett. {\bf 123}, 197702 (2019).

\bibitem{Lee2019}
J. Y. Lee, E. Khalaf, S. Liu, X. Liu, Z. Hao, P. Kim, and A. Vishwanath,
Theory of correlated insulating behaviour and spin-triplet superconductivity in twisted double bilayer graphene,
Nat. Commun. {\bf 10}, 5333 (2019).

\bibitem{Choi2019}
Y. W. Choi and H. J. Choi,
Intrinsic band gap and electrically tunable flat bands in twisted double bilayer graphene,
Phys. Rev. B {\bf 100}, 201402(R) (2019).

\bibitem{Liu2020}
X. Liu, Z. Hao, E. Khalaf, J. Y. Lee, Y. Ronen, H. Yoo, D. H. Najafabadi, K. Watanabe, T. Taniguchi, A. Vishwanath, and P. Kim,
Tunable spin-polarized correlated states in twisted double bilayer graphene,
Nature (London) {\bf 583}, 221 (2020).

\bibitem{JShin2022nearly}
J. Shin, B. L. Chittari, Y. Jang, H. Min, and J. Jung,
Nearly flat bands in twisted triple bilayer graphene,
Phys. Rev. B {\bf 105}, 245124 (2022).

\bibitem{Morell2013}
E. Su\'arez Morell, M. Pacheco, L. Chico, and L. Brey,
Electronic properties of twisted trilayer graphene,
Phys. Rev. B {\bf 87}, 125414 (2013).

\bibitem{Marton2020}
M. Szendrő, P. Süle, G. Dobrik and L. Tapasztó,
Ultra-flat twisted superlattices in 2D heterostructures,
npj Comput Mater {\bf 6}, 91 (2020).

\bibitem{Xu2021}
S. Xu, M. M. A. Ezzi, N. Balakrishnan, A. Garcia-Ruiz, B. Tsim, C. Mullan, J. Barrier, N. Xin, B. A. Piot, T. Taniguchi, K.Watanabe, A. Carvalho, A. Mishchenko, A. K. Geim, V. I. Fal’ko, S. Adam, A. H. C. Neto, K. S. Novoselov, and Y. Shi,
Tunable van Hove singularities and correlated states in twisted monolayer–bilayer graphene,
Nat. Phys. {\bf 17}, 619 (2021).

\bibitem{Chen2021}
S. Chen, M. He, Y.-H. Zhang, V. Hsieh, Z. Fei, K.Watanabe, T. Taniguchi, D. H. Cobden, X. Xu, C. R. Dean, and M. Yankowitz,
Electrically tunable correlated and topological states in twisted monolayer–bilayer graphene,
Nat. Phys. {\bf 17}, 374 (2021).

\bibitem{Polshyn2020}
H. Polshyn, J. Zhu, M. A. Kumar, Y. Zhang, F. Yang, C. L. Tschirhart, M. Serlin, K. Watanabe, T. Taniguchi, A. H. MacDonald, and A. F. Young,
Electrical switching of magnetic order in an orbital Chern insulator,
Nature (London) {\bf 588}, 66 (2020).

\bibitem{Park2020}
Y. Park, B. L. Chittari, and J. Jung,
Gate-tunable topological flat bands in twisted monolayer-bilayer graphene,
Phys. Rev. B {\bf 102}, 035411 (2020).

\bibitem{Lei2021}
C. Lei, L. Linhart, W. Qin, F. Libisch, and A. H. MacDonald,
Mirror symmetry breaking and lateral stacking shifts in twisted trilayer graphene,
Phys. Rev. B {\bf 104}, 035139 (2021).

\bibitem{Zewen2019}
Z. Wu, Z. Zhan, and S. Yuan,
Lattice relaxation, mirror symmetry and magnetic field effects on ultraflat bands in twisted trilayer graphene,
Sci. China Phys. Mech. Astron. {\bf 64}, 267811 (2021).

\bibitem{Xiao2019}
X. Li, F. Wu, and A. H. MacDonald,
Electronic structure of single-twist trilayer graphene,
arXiv:1907.12338.

\bibitem{Carr2020}
S. Carr, C. Li, Z. Zhu, E. Kaxiras, S. Sachdev, and A. Kruchkov,
Ultraheavy and Ultrarelativistic Dirac Quasiparticles in Sandwiched Graphenes,
Nano Lett. {\bf 20}, 3030 (2020).

\bibitem{MA202118}
Z. Ma, S. Li, Y.-W. Zheng, M.-M. Xiao, H. Jiang, J.-H. Gao, and X. Xie,
Topological flat bands in twisted trilayer graphene,
Science Bulletin {\bf 66}, 18 (2021).

\bibitem{Park2021}
J. M. Park, Y. Cao, K. Watanabe, T. Taniguchi, and P. Jarillo-Herrero,
Tunable strongly coupled superconductivity in magic-angle twisted trilayer graphene,
Nature (London) {\bf 590}, 249 (2021).

\bibitem{Hao2021}
Z. Hao, A. M. Zimmerman, P. Ledwith, E. Khalaf, D. H. Najafabadi, K. Watanabe, T. Taniguchi, A. Vishwanath, and P. Kim,
Electric field–tunable superconductivity in alternating-twist magic-angle trilayer graphene,
Science {\bf 371}, 6534 (2021).

\bibitem{Cao2021}
Y. Cao, J. M. Park, K. Watanabe, T. Taniguchi, and P. Jarillo-Herrero,
Pauli-limit violation and re-entrant superconductivity in moiré graphene,
Nature (London) {\bf 595}, 526 (2021).

\bibitem{Park2022}
J. M. Park, Y. Cao, L.-Q. Xia, S. Sun, K. Watanabe, T. Taniguchi, and P. Jarillo-Herrero,
Robust superconductivity in magic-angle multilayer graphene family,
Nat. Mater. {\bf 21}, 877 (2022).

\bibitem{Burg2022}
G. W. Burg, E. Khalaf, Y. Wang, K. Watanabe, T. Taniguchi, and E. Tutuc,
Emergence of correlations in alternating twist quadrilayer graphene,
Nat. Mater. {\bf 21}, 884 (2022).

\bibitem{Zhang2022}
Y. Zhang, R. Polski, C. Lewandowski, A. Thomson, Y. Peng, Y. Choi, H. Kim, K. Watanabe, T. Taniguchi, J. Alicea, F. von Oppen, G. Refael, and S. Nadj-Perge,
Promotion of superconductivity in magic-angle graphene multilayers,
Science {\bf 377}, 1538 (2022).

\bibitem{khalaf2019magic}
E. Khalaf, A. J. Kruchkov, G. Tarnopolsky, and A. Vishwanath,
Magic angle hierarchy in twisted graphene multilayers,
Phys. Rev. B {\bf 100}, 085109 (2019).

\bibitem{KShin2023}
K. Shin, Y. Jang, J. Shin, J. Jung, and H. Min,
Electronic structure of biased alternating-twist multilayer graphene,
Phys. Rev. B {\bf 107}, 245139 (2023).

\bibitem{JShin2021}
J. Shin, B. L. Chittari, and J. Jung,
Stacking and gate-tunable topological flat bands, gaps, and anisotropic strip patterns in twisted trilayer graphene,
Phys. Rev. B {\bf 104}, 045413 (2021).

\bibitem{Lopez-Bezanilla2020}
A. Lopez-Bezanilla and J. L. Lado,
Electrical band flattening, valley flux, and superconductivity in twisted trilayer graphene,
Phys. Rev. Res. {\bf 2}, 033357 (2020).

\bibitem{Phong2021}
V. T. Phong, P. A. Pantale\'{o}n, T. Cea, and F. Guinea,
Band structure and superconductivity in twisted trilayer graphene,
Phys. Rev. B {\bf 104}, L121116 (2021).

\bibitem{Dumitru2021}
D. C\ifmmode \u{a}\else \u{a}\fi{}lug\ifmmode \u{a}\else \u{a}\fi{}ru, F. Xie, Z.-D. Song, B. Lian, N. Regnault, and B. A. Bernevig,
Twisted symmetric trilayer graphene: Single-particle and many-body Hamiltonians and hidden nonlocal symmetries of trilayer moir\'e systems with and without displacement field,
Phys. Rev. B {\bf 103}, 195411 (2021).

\bibitem{Fang2021}
F. Xie, N. Regnault, D. C\ifmmode \u{a}\else \u{a}\fi{}lug\ifmmode \u{a}\else \u{a}\fi{}ru, B. A. Bernevig, and B. Lian,
Twisted symmetric trilayer graphene. II. Projected Hartree-Fock study,
Phys. Rev. B {\bf 104}, 115167 (2021).

\bibitem{Yu2023}
J. Yu, M. Xie, B. A. Bernevig, and S. Das Sarma,
Magic-angle twisted symmetric trilayer graphene as a topological heavy-fermion problem,
Phys. Rev. B {\bf 108}, 035129 (2023).

\bibitem{Kolar2023}
K. Kol\'a\ifmmode \check{r}\else \v{r}\fi{}, Y. Zhang, S. Nadj-Perge, F. von Oppen, and C. Lewandowski,
Electrostatic fate of $N$-layer moir\'e graphene,
Phys. Rev. B {\bf 108}, 195148 (2023).

\bibitem{Mora2019}
C. Mora, N. Regnault, and B. A. Bernevig,
Flatbands and Perfect Metal in Trilayer Moir\'e Graphene,
Phys. Rev. Lett. {\bf 123}, 026402 (2019).

\bibitem{Zhu2020}
Z. Zhu, S. Carr, D. Massatt, M. Luskin, and E. Kaxiras,
Twisted Trilayer Graphene: A Precisely Tunable Platform for Correlated Electrons,
Phys. Rev. Lett. {\bf 125}, 116404 (2020).

\bibitem{Zhang2021}
X. Zhang, K.-T. Tsai, Z. Zhu, W. Ren, Y. Luo, S. Carr, M. Luskin, E. Kaxiras, and K. Wang,
Correlated Insulating States and Transport Signature of Superconductivity in Twisted Trilayer Graphene Superlattices,
Phys. Rev. Lett. {\bf 127}, 166802 (2021).

\bibitem{Mao2023}
Y. Mao, D. Guerci, and C. Mora,
Supermoir\'e low-energy effective theory of twisted trilayer graphene,
Phys. Rev. B {\bf 107}, 125423 (2023).

\bibitem{Devakul2023}
T. Devakul, P. J. Ledwith, L.-Q. Xia, A. Uri, S. C. de la Barrera, P. Jarillo-Herrero, and L. Fu,
Magic-angle helical trilayer graphene,
Science Advances {\bf 9}, eadi6063 (2023).

\bibitem{Popov2023}
F. K. Popov and G. Tarnopolsky,
Magic angles in equal-twist trilayer graphene,
Phys. Rev. B {\bf 108}, L081124 (2023).

\bibitem{Guerci2023}
D. Guerci, Y. Mao, and C. Mora,
Chern mosaic and ideal flat bands in equal-twist trilayer graphene,
Phys. Rev. Res. {\bf 6}, L022025 (2024).

\bibitem{Nakatsuji2023}
N. Nakatsuji, T. Kawakami, and M. Koshino,
Multiscale Lattice Relaxation in General Twisted Trilayer Graphenes,
Phys. Rev. X {\bf 13}, 041007 (2023).

\bibitem{Louk2019}
L. Rademaker, D. A. Abanin, and P. Mellado,
Charge smoothening and band flattening due to Hartree corrections in twisted bilayer graphene,
Phys. Rev. B {\bf 100}, 205114 (2019).

\bibitem{Goodwin2020}
Z. A. H. Goodwin, V. Vitale, X. Liang, A. A. Mostofi, and J. Lischner,
Hartree theory calculations of quasiparticle properties in twisted bilayer graphene,
Electron. Struct. {\bf 2}, 034001 (2020).

\bibitem{Guinea2018}
F. Guinea and N. R. Walet,
Electrostatic effects, band distortions, and superconductivity in twisted graphene bilayers,
Proc. Natl. Acad. Sci. U.S.A. {\bf 115}, 13174 (2018).

\bibitem{Cea2019}
T. Cea, N. R. Walet, and F. Guinea,
Electronic band structure and pinning of Fermi energy to Van Hove singularities in twisted bilayer graphene: A self-consistent approach,
Phys. Rev. B {\bf 100}, 205113 (2019).

\bibitem{Cea2020}
T. Cea, and F. Guinea,
Band structure and insulating states driven by Coulomb interaction in twisted bilayer graphene,
Phys. Rev. B {\bf 102}, 045107 (2020).

\bibitem{Cea2022}
T. Cea, P. A. Pantale{\'o}n, N. R. Walet, and F. Guinea,
Electrostatic interactions in twisted bilayer graphene,
Nano Mater. Sci. {\bf 4}, 27 (2022).

\bibitem{Novelli2020}
P. Novelli, I. Torre, F. H. L. Koppens, F. Taddei, and M. Polini,
Optical and plasmonic properties of twisted bilayer graphene: Impact of interlayer tunneling asymmetry and ground-state charge inhomogeneity,
Phys. Rev. B {\bf 102}, 125403 (2020).

\bibitem{Ding2022}
C. Ding, X. Zhang, H. Gao, X. Ma, Y. Li, and M. Zhao,
Role of electron-electron interaction in the plasmon modes of twisted bilayer graphene,
Phys. Rev. B {\bf 106}, 155402 (2022).

\bibitem{Lewandowski2021}
C. Lewandowski, S. Nadj-Perge, and D. Chowdhury,
Does filling-dependent band renormalization aid pairing in twisted bilayer graphene?,
npj Quantum Mater. {\bf 6}, 82 (2021).

\bibitem{Choi2021}
Y. Choi, H. Kim, C. Lewandowski, Y. Peng, A. Thomson, R. Polski, Y. Zhang, K. Watanabe, T. Taniguchi, J. Alicea, and S. Nadj-Perge,
Interaction-driven band flattening and correlated phases in twisted bilayer graphene,
Nat. Phys. {\bf 17}, 1375 (2021).

\bibitem{Zhang2020}
Y. Zhang, K. Jiang, Z. Wang, and F. Zhang,
Correlated insulating phases of twisted bilayer graphene at commensurate filling fractions: A Hartree-Fock study,
Phys. Rev. B {\bf 102}, 035136 (2020).

\bibitem{Ming2020}
M. Xie and A. H. MacDonald,
Nature of the Correlated Insulator States in Twisted Bilayer Graphene,
Phys. Rev. Lett. {\bf 124}, 097601 (2020).

\bibitem{Ming2021}
M. Xie and A. H. MacDonald,
Weak-Field Hall Resistivity and Spin-Valley Flavor Symmetry Breaking in Magic-Angle Twisted Bilayer Graphene,
Phys. Rev. Lett. {\bf 127}, 196401 (2021).

\bibitem{Wei2023}
W. Qin, B. Zou, and A. H. MacDonald,
Critical magnetic fields and electron pairing in magic-angle twisted bilayer graphene,
Phys. Rev. B {\bf 107}, 024509 (2023).

\bibitem{Jung2014}
J. Jung, A. Raoux, Z. Qiao, and A. H. MacDonald,
${\it Ab}$ ${\it initio}$ theory of moir{\'{e}} superlattice bands in layered two-dimensional materials,
Phys. Rev. B {\bf 89}, 205414 (2014).

\bibitem{Bistritzer2010b}
R. Bistritzer and A. H. MacDonald,
Moir\'{e} bands in twisted double-layer graphene,
Proc. Natl. Acad. Sci. U.S.A. {\bf 108}, 12233 (2011).

\bibitem{Jung2013}
J. Jung and A. H. MacDonald,
Gapped broken symmetry states in ABC-stacked trilayer graphene,
Phys. Rev. B {\bf 88}, 075408 (2013).

\bibitem{Xiao2010}
D. Xiao, M.-C. Chang, and Q. Niu,
Berry phase effects on electronic properties,
Rev. Mod. Phys. {\bf 82}, 1959 (2010).

\bibitem{Stauber2013}
T. Stauber, P. San-Jose, and L. Brey,
Optical conductivity, Drude weight and plasmons in twisted graphene bilayers,
New Journal of Physics {\bf 15}, 113050 (2013).

\bibitem{Min2009}
H. Min and A. H. MacDonald,
Origin of Universal Optical Conductivity and Optical Stacking Sequence Identification in Multilayer Graphene,
Phys. Rev. Lett. {\bf 103}, 067402 (2009).

\bibitem{Laissardiere2010}
G. Trambly de Laissardière, D. Mayou, and L. Magaud,
Localization of Dirac Electrons in Rotated Graphene Bilayers,
Nano Lett. {\bf 10}, 804 (2010).

\bibitem{Uchida2014}
K. Uchida, S. Furuya, J.-I. Iwata, and A. Oshiyama,
Atomic corrugation and electron localization due to Moir\'e patterns in twisted bilayer graphenes,
Phys. Rev. B {\bf 90}, 155451 (2014).

\bibitem{Koshino2018}
M. Koshino, N. F. Q. Yuan, T. Koretsune, M. Ochi, K. Kuroki, and L. Fu,
Maximally Localized Wannier Orbitals and the Extended Hubbard Model for Twisted Bilayer Graphene,
Phys. Rev. X {\bf 8}, 031087 (2018).

\bibitem{YChoi2019}
Y. Choi, J. Kemmer, Y. Peng, A. Thomson, H. Arora, R. Polski, Y. Zhang, H. Ren, J. Alicea, G. Refael, F. von Oppen, K. Watanabe, T. Taniguchi, and S. Nadj-Perge,
Electronic correlations in twisted bilayer graphene near the magic angle,
Nat. Phys. {\bf 15}, 1174 (2019).

\bibitem{Bernevig2021}
B. A. Bernevig, Z.-D. Song, N. Regnault, and B. Lian,
Twisted bilayer graphene. III. Interacting Hamiltonian and exact symmetries,
Phys. Rev. B {\bf 103}, 205413 (2021).


\end{thebibliography}
\end{document}